\documentclass[twocolumn,superscriptaddress,longbibliography,aps,prb,amsmath,preprintnumbers]{revtex4-1}
\usepackage{tikz}
\usepackage{hyperref}
\usepackage{epsfig}

\begin{document}
\title{How long does it take to form the Andreev quasiparticles?}

\author{R. Taranko}
\author{T.\ Doma\'nski}
\email{doman@kft.umcs.lublin.pl}
\affiliation{Institute of Physics, M.\ Curie Sk\l odowska University, 20-031 Lublin, Poland}

\date{\today}

\begin{abstract}
We study transient effects in a setup, where the quantum dot (QD) is abruptly sandwiched
between the metallic and superconducting leads. Focusing on the proximity-induced electron
pairing, manifested by the in-gap bound states, we determine  characteristic time-scale
needed for these quasiparticles to develop. In particular, we derive analytic  expressions
for (i) charge occupancy of the QD, (ii) amplitude of the induced electron pairing, and
(iii) the transient currents under equilibrium and nonequilibrium conditions. We also
investigate the correlation effects within the Hartree-Fock-Bogolubov approximation,
revealing a competition between the Coulomb interactions and electron pairing.
\end{abstract}



\maketitle

\section{Introduction\label{sec:intro}}

Quantum impurity attached to a superconducting bulk material absorbs the Cooper pairs, developing the quasiparticle
states in its subgap spectrum $|\omega|\leq\Delta$, where $\Delta$ is the energy gap of superconducting reservoir
\cite{balatsky.vekhter.06,Rodero-11}. These bound Andreev (or Yu-Shiba-Rusinov) states have been observed in numerous
STM studies, using  impurities deposited on superconducting substrates \cite{STM-1} and in tunneling experiments via
quantum dots arranged in the Josephson \cite{Josephson-1}, Andreev \cite{Andreev-1} and more complex (multi-terminal)
configurations \cite{multiterminal-1,Baumgartner-17}. Since measurements can be nowadays done with state-of-art
precision probing the time-resolved properties, we address this issue here and determine some characteristic 
temporal scales of the in-gap quasiparticles.

Any abrupt change of the model parameters ({\em quantum quench}) is usually followed by a time-dependent
thermalization of the many-body system , where continuum states play a prominent role \cite{LevyYeyati-2017}.
Dynamics of these processes has been recently explored in the solid state and nanoscopic physics \cite{Gogolin-2015}.
From a practical point of view,  especially useful could be nanoscopic
heterostructures with the correlated quantum dot (QD) embedded between  external (metallic, ferromagnetic or
superconducting) leads which enable measurements of the transport properties under tunable nonequilibrium conditions
\cite{Bidzhiev-2017}.

Transport phenomena through QD coupled between the normal or superconducting leads have so far 
explored mainly in the static cases. Since novel experimental methods allow to study the QDs subjected to 
voltage pulses or abrupt changes of the system parameters, it would be very desirable to calculate 
the time-dependent currents and their conductances. In particular, one can ask the question: 
{\em how fast does the QD respond to an instantaneous perturbation}. For this purpose  analytical 
estimation of the transient oscillations and long-time (asymptotic) behaviour of the measurable
quantities would be very useful. Some early theoretical works have investigated time-dependent 
transport via QD between the normal and superconducting leads
\cite{Sun-2000,Zhao-2001,Wei-2002,Cao-2015}, however, analytic results are hardly 
available. As regards the QD coupled to both normal leads, the transient current and charge occupancy 
have been determined for abrupt voltage pulses or after an instantaneous switching of constituent 
parts of the system~\cite{A1,A2,A3,A4,A5,A6,A7,A8,A9,A10,A11,A12,A13}.

Time-resolved techniques could provide an insight into the many-body effects. For instance, the 
pump-and-probe experiments \cite{Orenstein-15} and the time-resolved ARPES \cite{t-ARPES} have 
determined life-time of the Bogoliubov quasiparticles in the high temperature superconductors. 
Transient effects have been investigated in nanoscopic systems, considering mainly the quantum 
dots hybridized with the conducting (metallic) leads. There has been studied the 
time-scale needed for the Kondo peak to develop at the Fermi energy \cite{Nordlander-99},
dynamical correlations in electronic transport via the quantum dots \cite{Michalek-09}, or 
oscillatory behavior in the charge transport through the molecular junctions \cite{Yeyati-15}.

Dynamical phenomena of the quantum dots attached to superconducting bulk reservoirs have been studied much
less intensively. There has been analyzed: photon-assisted Andreev tunneling
\cite{Sun-99}, response time on a step-like pulse \cite{Xing-07}, temporal dependence of the multiple Andreev
reflections \cite{Stefanucci-10}, time-dependent sequential tunneling \cite{Konig-12}, transient effects caused by an
oscillating level \cite{Komnik-13}, time-dependent bias \cite{Pototzky-14}, the waiting time distributions in
nonequilibrium transport \cite{Governale-13,Michalek-17}, the short-time counting statistics \cite{Konig-16},
metastable configurations of the Andreev bound states in a phase-biased Josephson junction
\cite{LevyYeyati-2016,LevyYeyati-2017} etc.  None of these studies, however, addressed the time-scale
typical for development of the subgap quasiparticle states in a setup, comprising the quantum dot (QD)  coupled
to the normal lead (N) on one side and to the isotropic ($s$-wave) superconductor (S) on the other side. Our present
study reveals, that a continuous electronic spectrum of the metallic lead enables a relaxation of the Andreev states,
whereas the superconducting electrode induces the (damped) quantum oscillations with a period sensitive to the energies of the in-gap quasiparticles. In what follows we evaluate the time-scale at which such Andreev 
quasiparticle start to form and another one, when they are finally established.

The paper is organized as follows. In Sec.\ \ref{sec:micro-model} we introduce
the microscopic model and discuss the method for the time-dependent
phenomena. Sec.\ \ref{sec:N-QD-S} presents  analytical results
for the uncorrelated quantum dot, such as: (i) charge occupancy,
(ii) complex order parameter, and (iii) charge current for the unbiased and
biased heterojunction. In Sec.\ \ref{sec:Correlations} we discuss
the correlation effects and finally in Sec.\ \ref{sec:Summary} we summarize
the main results.

\section{Microscopic model\label{sec:micro-model}}
%
\begin{figure}[t]
\includegraphics[width=0.7\columnwidth]{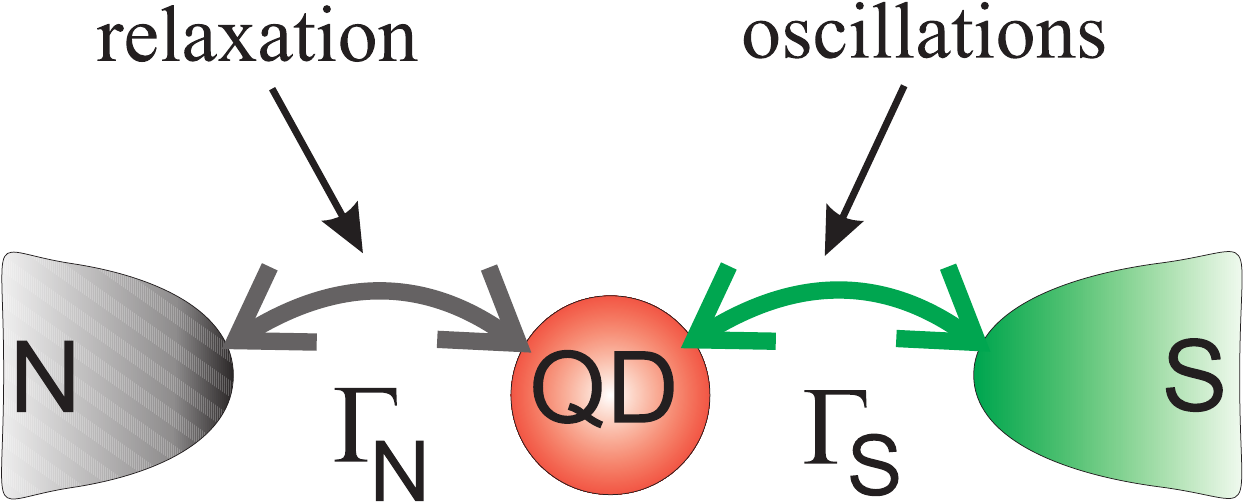}
\caption{Schematics of the setup, comprising the quantum dot (QD) coupled 
to the normal (N) and superconducting (S) electrodes. Sudden coupling to 
the continuum states triggers the relaxation processes, whereas superconductr
induces the in-gap bound states giving rise to quantum oscillations.} 
\label{schem}
\end{figure}
For description of the N-QD-S heterostructure (see Fig.~\ref{schem}) we use the single impurity Anderson Hamiltonian
\begin{eqnarray}
\hat{H} =  \sum_{\sigma} \varepsilon_{\sigma} \hat{d}^{\dagger}_{\sigma}
\hat{d}_{\sigma}  +  U \; \hat{n}_{\uparrow} \hat{n}_{\downarrow}  +
\sum_{\beta} \left( \hat{H}_{\beta} + \hat{V}_{\beta - QD} \right)
\label{model}
\end{eqnarray}
where $\beta$ refers to the normal ($N$) and superconducting ($S$) electrodes, respectively. As usually
$\hat{d}_{\sigma}$ ($\hat{d}^{\dagger}_{\sigma}$) is the annihilation (creation) operator for the quantum dot (QD)
electron with spin $\sigma$ and energy $\varepsilon_{\sigma}$. Potential of the Coulomb repulsion between the opposite
spin electrons is denoted by $U$. We treat the external metallic lead as free fermion gas $\hat{H}_{N} \!=\! \sum_{{\bf
k},\sigma} \varepsilon_{\bf k} \hat{c}_{{\bf k} \sigma}^{\dagger} \hat{c}_{{\bf k} \sigma}$, and describe the isotropic
superconductor  by the BCS model $\hat{H}_{S} \!=\!\sum_{{\bf q},\sigma}  \varepsilon_{{\bf q}} \hat{c}_{{\bf q}
\sigma}^{\dagger}  \hat{c}_{{\bf q} \sigma} \!-\! \sum_{\bf q} \Delta  \left( \hat{c}_{{\bf q} \uparrow} ^{\dagger}
\hat{c}_{-{\bf q} \downarrow}^{\dagger} + \hat{c} _{-{\bf q} \downarrow} \hat{c}_{{\bf q} \uparrow }\right)$, where
$\varepsilon_{{\bf k} ({\bf q})}$ is the energy measured from the chemical potential $\mu_{N(S)}$, and $\Delta$ denotes
the superconducting energy gap. Hybridization between the QD electrons and the metallic lead is given by $ \hat{V}_{N -
QD} = \sum_{{\bf k},\sigma} \left( V_{{\bf k}} \; \hat{d}_{\sigma}^{\dagger}  \hat{c}_{{\bf k} \sigma} + \mbox{\rm
h.c.} \right)$ and $\hat{V}_{S - QD}$ can be expressed by interchanging ${\bf k} \leftrightarrow {\bf q}$.

Since our study refers to the subgap quasiparticle states,
we assume the constant couplings $\Gamma_{N (S)}=2\pi
\sum_{{\bf k}({\bf q})} |V_{{\bf k}({\bf q})}|^2 \;
\delta(\omega \!-\! \varepsilon_{{\bf k}({\bf q})})$. In the deep
subgap regime $|\omega|\ll\Delta$ (so called, superconducting atomic limit)
the coupling $\Gamma_{S}/2$ can be regarded as a qualitative measure of
the induced pairing potential, whereas $\Gamma_{N}$ controls the inverse
life-time of the in-gap quasiparticles. As we shall see, both these
couplings play important (though quite different) role in transient phenomena.


We assume that all three constituents of the N-QD-S heterostructure are
disconnected from each other until $t\leq 0$. Let us impose the external
(N, S) reservoirs to be suddenly coupled to the quantum dot
\begin{eqnarray}
V_{{\bf k}({\bf q})}(t) =
\left\{ \begin{array}{cc}
0 & \hspace{0.5cm} \mbox{\rm for } t \leq 0  \\
V_{{\bf k}({\bf q})}  & \hspace{0.5cm} \mbox{\rm for } t > 0 ,
\end{array} \right.
\label{abrupt_coupling}
\end{eqnarray}
inducing the transient effects. Later on, we shall relax this assumption. Our problem resembles the
Wiener-Hopf method \cite{Janis-1997} applied earlier in the studies of X-ray absorption and emission of metals
\cite{DeDominicis-1969}.

In what follows, we explore the time-dependence of physical observables $\hat{O}$, based on the Heisenberg equation of
motion $i\hbar \frac{d}{dt}\hat{O} = \left[ \hat{O},\hat{H} \right]$. In particular, we shall investigate expectation
values of the QD occupancy $\langle \hat{d}_{\sigma}^{\dagger}(t)\hat{d}_{\sigma}(t)\rangle$, the induced on-dot
pairing $\langle \hat{d}_{\downarrow}(t) \hat{d}_{\uparrow}(t)\rangle$, and the transient charge currents flowing
between the QD and external electrodes (both under equilibrium and nonequilibrium condictions).


Our strategy is based on the following three steps. First, we formulate the differential equations of motion for the
annihilation $\hat{d}_{\sigma}(t)$ and creation $\hat{d}_{\sigma}^{\dagger}(t)$ operators of QD and similar ones for
the mobile electrons  $\hat{c}_{{\bf k}({\bf q})\sigma}(t)$ and $\hat{c}_{{\bf k}({\bf q})\sigma}^{(\dagger)}(t)$,
respectively. Next, we solve them using the Laplace transformations, e.g. for $\hat{d}_{\sigma}(t)$ we denote
\begin{eqnarray}
\hat{d}_{\sigma}(s) = \int_{0}^{\infty} e^{-st}\hat{d}_{\sigma}(t)dt \equiv {\cal{L}}\left\{
\hat{d}_{\sigma}(t)\right\}(s) . \label{Laplace_transform}
\end{eqnarray}
For the uncorrelated QD the analytical expressions for  $\hat{d}_{\sigma}(s)$ and $\hat{d}_{\sigma}^{\dagger}(s)$ can
be obtained (see Appendix \ref{sec:Laplace}). Finally, using the corresponding inverse Laplace transforms, we compute
the time-dependent expectation values of a the QD occupancy, the QD pair amplitude and currents flowing between QD and
both leads. For example QD occupancy $n_{\sigma}(t) \equiv \langle
\hat{d}_{\sigma}^{\dagger}(t)\hat{d}_{\sigma}(t)\rangle$ is given  by
\begin{equation}
n_{\sigma}(t) = \left< {\cal{L}}^{-1} \!\left\{
\hat{d}^{\dagger}_{\sigma}(s) \right\}\!(t) \;{\cal{L}}^{-1}
\!\left\{ \hat{d}_{\sigma}(s) \right\}\!(t) \right> ,
\label{QD_occupancy}
\end{equation}
where ${\cal{L}}^{-1} \left\{ \hat{d}_{\sigma}(s)\right\}(t)$ stands for the inverse Laplace transform of
$\hat{d}_{\sigma}(s)$.

In our calculations we make use of the wide-band limit approximation ($\Gamma_{\beta}=$const) and set
$e=\hbar=k_{B}\equiv 1$, so that all energies, currents and time are expressed in units of $\Gamma_{S}$,
$e\Gamma_{S}/\hbar$ and $\hbar/\Gamma_{S}$, respectively. We also treat the chemical potential $\mu_{S}=0$ 
as a convenient reference energy point and perform the calculations for zero temperature.  For  
experimentally available value $\Gamma_S \sim 200 \mu eV$ \cite{Deacon-10,Pillet-13,Eichler2007}, 
the typical period of transient oscillations would be $\sim 3.3 psec$. 

\section{Uncorrelated QD case\label{sec:N-QD-S}}

We start by addressing the transient effects of the uncorrelated quantum dot ($U=0$), 
focusing on the superconducting atomic limit ($\Delta=\infty$) for which analytical 
expressions can be obtained. More general cosiderations are presented in Appendix \ref{A}.

\subsection{Time-dependent QD charge \label{sec:N-QD-S_occupancy}}

Let us inspect the time-dependent occupancy $n_{\sigma}(t)$ driven by an abrupt coupling  
of the QD to both external leads.  This quantity, defined in Eq.~(\ref{QD_occupancy}), 
can be determined explicitely for arbitratry $\Delta$ (derivation is presented in
Appendix \ref{A}). Here we shall consider the formula (\ref{A.14}) simplified for 
the superconducting atomic limit
\begin{widetext}
\begin{eqnarray}
n_{\uparrow}(t) &=&  e^{-\Gamma_{N} t} \left\{ n_{\uparrow}(0) + \left[ 1 -
n_{\uparrow}(0) - n_{\downarrow}(0) \right] \; \mbox{\rm sin}^{2}
\left( \frac{\sqrt{\delta}}{2}\;t\right) \; \frac{\Gamma_{S}^{2}}{\delta}  \right\}
\label{QD_for_NQDS}  \\
& + &  \frac{\Gamma_{N}}{2\pi} \int_{-\infty}^{\infty} \hspace{-0.3cm} d\omega \;\; f_N(\omega) \; {\cal{L}}^{-1} \left\{
\frac{s+i\varepsilon_{-\sigma}+\Gamma_{N}/2} {(s-s_{1})(s-s_{2})(s-i\omega)} \right\} \!(t) \;\;
{\cal{L}}^{-1} \! \left\{ \frac{s-i\varepsilon_{-\sigma}+\Gamma_{N}/2}
{(s-s_{3})(s-s_{4})(s+i\omega)}  \right\} \!(t)  \nonumber \\
& + &  \frac{\Gamma_{N}}{2\pi}  \int_{-\infty}^{\infty} \hspace{-0.3cm} d\omega
\;\; \left[ 1 - f_N(\omega) \right] \; {\cal{L}}^{-1} \! \left\{ \frac{\Gamma_{S}/2}
{(s-s_{1})(s-s_{2})(s+i\omega )}  \right\} \!(t) \;\; {\cal{L}}^{-1} \! \left\{ \frac{\Gamma_{S}/2}
{(s-s_{3})(s-s_{4})(s-i\omega )}  \right\} \!(t)   , \nonumber
\end{eqnarray}
\end{widetext}
where $f_N(\omega)$ is the Fermi-Dirac distribution function of the normal lead and parameters 
$s_1, s_2, s_3, s_4$ and $\delta$ are presented in Eq.~(\ref{A.12}). The  occupancy 
$n_{\downarrow}(t)$ can be obtained from the same expression (\ref{QD_for_NQDS})
upon replacing the set $(s_{1},s_{2},s_{3},s_{4})$ by $(s_{3},s_{4},s_{1},s_{2})$. 
Expressions given in the second and third lines of Eq.~(\ref{QD_for_NQDS}) could be 
presented in the compact analytical form in the case $\varepsilon_{\sigma}=0$ (see 
Eqs.~(\ref{A.17}-\ref{A.19})). Otherwise they are rather lengthy (even though accessible), 
therefore we skip them.

Another  simplification of Eq.~(\ref{QD_for_NQDS}) is possible upon neglecting the normal lead
($\Gamma_{N}=0$). QD occupancy is then characterized by  non-vanishing quantum oscillations
\begin{eqnarray}
n_{\sigma}(t)=  n_{\sigma}(0) + \left[ 1 - n_{\sigma}(0)-n_{-\sigma}(0) \right] \; \mbox{\rm sin}^{2} \! \left(
\frac{\sqrt{\delta}}{2}\;t \right) \; \frac{\Gamma_{S}^{2}}{\delta} .\nonumber\\ \label{simple_QD_occup}
\end{eqnarray}
For $\varepsilon_{\sigma}\!=\!0$ Eq.\ (\ref{simple_QD_occup}) reduces to
\begin{eqnarray}
n_{\sigma}(t)= \cos^{2} \! \left( \frac{{\Gamma_S}}{2}\;t \right) n_{\sigma}(0) + \sin^{2} \! \left(
\frac{{\Gamma_S}}{2}\;t \right) \left[ 1\!-\! n_{-{\sigma}}(0) \right] , \nonumber\\
\label{very_simple_QD_occup}
\end{eqnarray}
implying the period of transient oscillations $T=2\pi/\Gamma_{S}$, except of the initial 
conditions $n_{\sigma}(0)=1$ and $n_{-{\sigma}}(0)=0$ when the QD occupancy is preserved.

\begin{figure}[b]
\includegraphics[width=0.95\columnwidth]{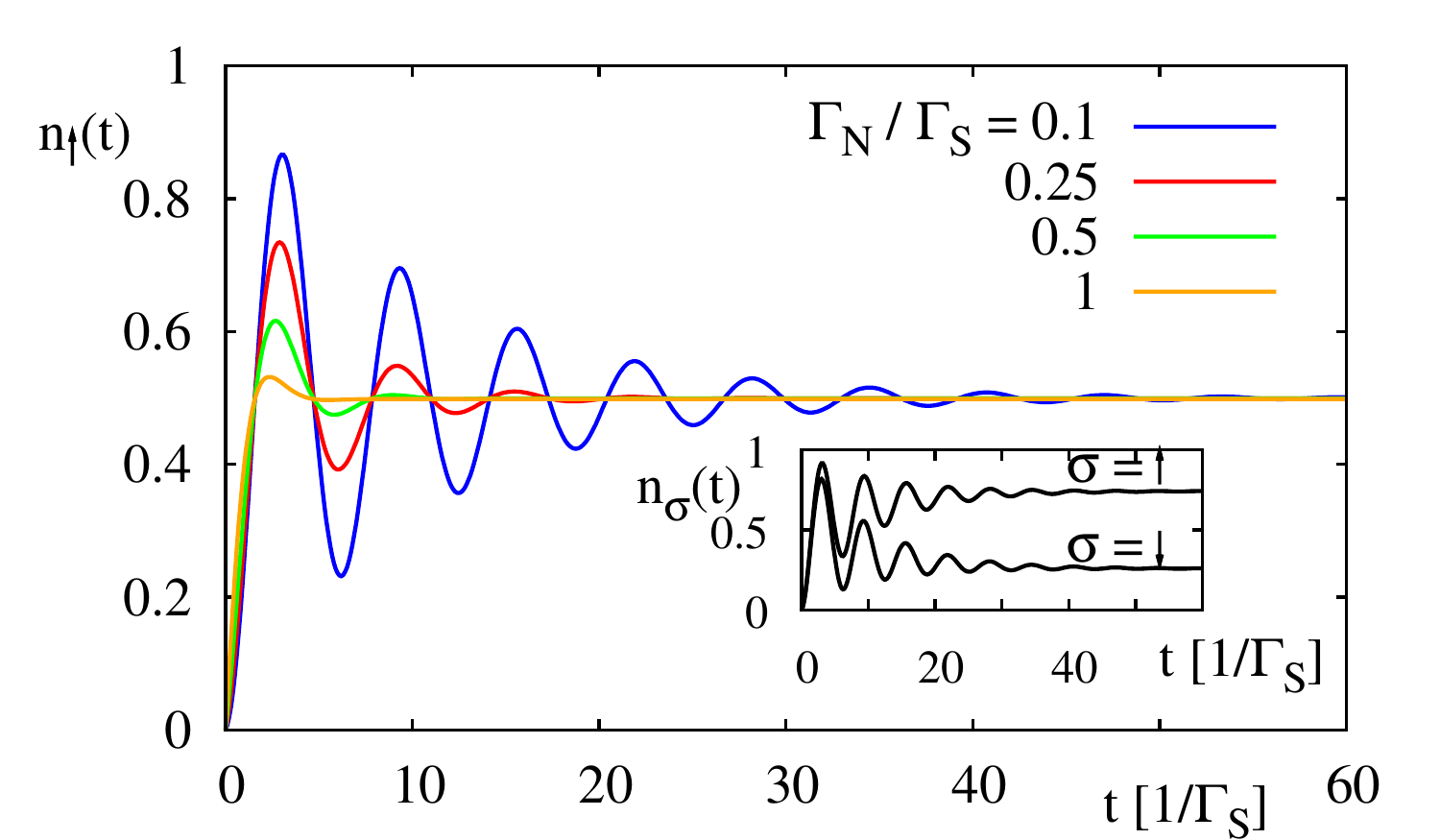}
\caption{Time-dependent occupancy $n_{\uparrow}(t)=n_{\downarrow}(t)$ obtained for
$\varepsilon_{\sigma}=0$, assuming the initial occupancy $n_{\uparrow}(0)=0=n_{\downarrow}(0)$ 
in absence of external voltage ($\mu_{N}=\mu_{S}=0$). Different lines correspond to various 
ratios $\Gamma_{N}/\Gamma_{S}$, indicated in the legend. Inset shows the QD occupancies
$n_{\sigma}(t)$ for the finite Zeeman splitting $\varepsilon_{\downarrow}-\varepsilon_{\uparrow}
=\Gamma_{S}$, assuming $\Gamma_{N}/\Gamma_{S}=0.1$.
}
\label{figure2}
\end{figure}

The formula (\ref{very_simple_QD_occup}), obtained in the case $\Gamma_{N}=0$, resembles 
the Rabi oscillations of a typical two-level quantum system. Indeed, the proximitized QD 
is fully equivalent to such scenario. To prove it, let us consider the effective Hamiltonian 
$\hat{H}=\sum_{\sigma}\varepsilon_{\sigma}\hat{n}_{\sigma}+ \frac{\Gamma_{S}}{2} \left(
\hat{d}^{\dagger}_{\uparrow}\hat{d}^{\dagger}_{\downarrow} + \mbox{\rm h.c.}\right)$, 
assuming that at $t=0$ the QD  is empty $n_{\uparrow}(0)=0=n_{\downarrow}(0)$. For arbitrary 
time $t>0$ we can calculate the probability $P(t)$ of finding the QD  in the doubly occupied 
configuration $n_{\uparrow}(t)=1=n_{\downarrow}(t)$ within the standard treatment of a two-level 
system \cite{Rabbi}. This probability is given by
\begin{equation}
P(t)= \frac{\Gamma_{S}^{2}}{(E_{1}-E_{2})^{2}+\Gamma_{S}^{2}}
\mbox{\rm sin}^{2}\left( \frac{t}{2}\sqrt{(E_{1}-E_{2})^{2}+\Gamma_{S}^{2}}\right) ,
\label{Rabbi_formula}
\end{equation}
where $E_{1}=0$ and $E_{2}=\varepsilon_{\uparrow}+\varepsilon_{\downarrow}$ are the energies of empty and doubly
occupied configurations, respectively. This result exactly reproduces our expression (\ref{very_simple_QD_occup}). 

\begin{figure}[t]
\includegraphics[width=0.95\columnwidth]{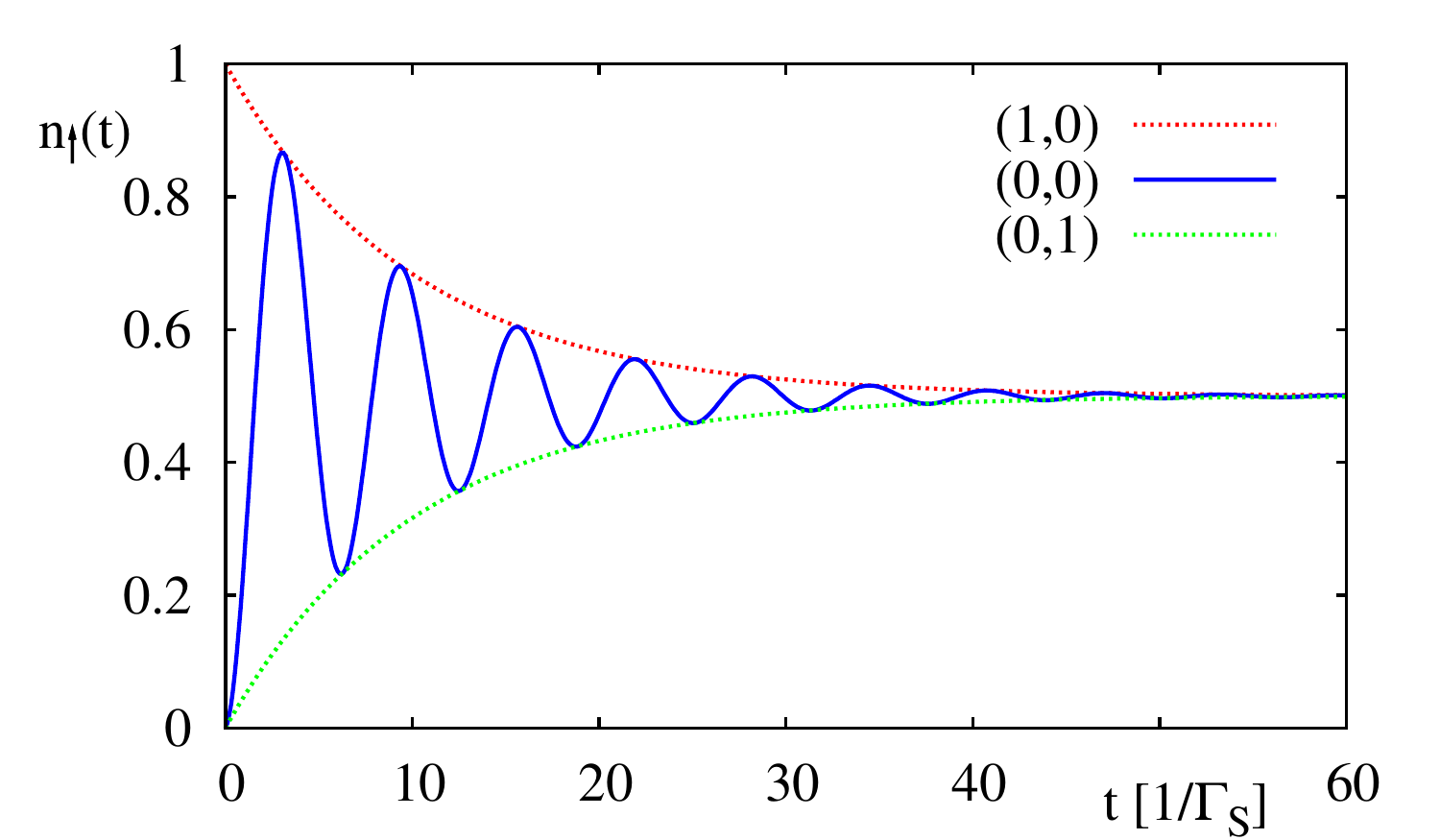}
\caption{
The time-dependent QD occupancy $n_{\uparrow}(t)$ obtained
in absence of external voltage  for $\varepsilon_{\sigma}=0$, $\Gamma_{N}=0.1\Gamma_{S}$.
Different curves refer to various initial occupancies $\left( n_{\uparrow}(0),
n_{\downarrow}(0)\right)$ indicated in the legend.}
\label{figure1}
\end{figure}

For the QD suddenly coupled to both the normal and superconducting leads ($\Gamma_{N,S}\neq 0$) 
such oscillations become damped (see Fig.\ \ref{figure2}). This effect comes partly from
the exponential factor $\exp(-\Gamma_{N}t)$ appearing in front of the first term in Eq.\ (\ref{QD_for_NQDS}) and partly
from the second and third contributions. This can be illustrated, by considering the case $\varepsilon_{\sigma}=0$,
 $\mu_N=0$, for which Eq.\ (\ref{QD_for_NQDS}) implies
\begin{eqnarray} \label{damped_oscil}
n_{\sigma}(t)&=& e^{-\Gamma_{N}t} \left\{ \cos^{2} \! \left( \frac{\Gamma_S}{2} t \right) n_{\sigma}(0)  \right.
 \\ &+& \left. \sin^{2} \! \left( \frac{\Gamma_S}{2} t \right) \left[ 1\!-\! n_{-{\sigma}}(0) \right] \right\}
+ \frac{1}{2} \left( 1 - e^{-\Gamma_{N}t} \right) .\nonumber
\end{eqnarray}
Under such circumstances, the QD occupancy approaches asymptotically a half-filling,
$\lim_{t\rightarrow\infty}n_{\sigma}(t) = \frac{1}{2}$. Fig.\ \ref{figure2} displays $n_{\uparrow}(t)$ obtained in
absence of external voltage for several values of $\Gamma_{N}$, assuming $\varepsilon_{\sigma}=0$ and $n_{\sigma}(0)=0$
for both spins. The quantum oscillations occur with a period $2\pi/\Gamma_{S}$ and their damping is governed by the
envelope function $e^{-\Gamma_{N}t}$ indicating, that a continuous spectrum of the metallic lead is responsible
for the relaxation processes. For a weak enough coupling $\Gamma_{N}$ these oscillations could indirectly probe 
the dynamical transitions between the subgap bound states, as recently emphasized by J.\ Gramich {\em et al}
[\onlinecite{Baumgartner-17}].

Fig.\ \ref{figure1} shows the QD occupancies obtained for several initial conditions, assuming  $\mu_{N}=\mu_{S}=0$
and $\varepsilon_{\sigma}=0$. The case $n_{\downarrow}(0)=0=n_{\uparrow}(0)$ allows quantum oscillations between
two eigenstates of the proximitized QD which are damped due to coupling to the normal lead (see Fig.\ \ref{figure2}).
For the initial condition $n_{\sigma}(0)=1$,  $n_{-\sigma}(0)=0$, the transient effects are completely different. The
first term in Eq.~(\ref{damped_oscil}) for $(n_{\uparrow}(0), n_{\downarrow}(0))=(1,0)$ or $(0,1)$ equals $e^{-\Gamma_N
t}$ or vanishes and together with the last term they yield $\frac{1}{2} (1-e^{-\Gamma_N t})$ - see the upper
curve in Fig.~\ref{figure1} or $\frac{1}{2} (1+e^{-\Gamma_N t})$ - the lower curve, respectively.  This stems from the
fact that proximity-induced pairing affects only the empty and doubly occupied configurations and it is inefficient in the case
considered here. In consequence the quantum oscillations are absent and the QD occupancy exponentially evolves towards
a halfilling. Let us also remark, that for $\Gamma_{S}=0$ Eq.\
(\ref{QD_for_NQDS}) simplifies to the standard formula obtained by the non-equilibrium Green's function method
\cite{Jauho-1994} (see Eq.~\ref{A.20}).

\subsection{Development of the proximity effect \label{sec:N-QD-S_pairing}}

Occupancy of the QD only indirectly tells us about emergence of the subgap bound states. To get some insight into the
superconducting proximity effect we shall study here the time evolution of the order parameter $\chi(t) = \left<
\hat{d}_{\downarrow}(t)\hat{d}_{\uparrow}(t)\right>$. The general formula is explicitly given by Eq.\ (\ref{A.23}).
Expressing its first two terms (which depend on the initial QD occupancy) the pair correlation
function can be written as 
\begin{eqnarray}
\chi(t) &=&    \left[ (\varepsilon_{\uparrow} +\varepsilon_{\downarrow}) \left( 1 - \mbox{\rm
cos}\left(\sqrt{\delta}\;t\right) \right) +i\sqrt{\delta} \mbox{\rm sin}\left( \sqrt{\delta}\;t\right) \right]
\nonumber
 \\ &\times & e^{-\Gamma_{N}t} \Gamma_{S} \frac{ (n_{\uparrow}(0)+n_{\downarrow}(0) - 1) }{2\delta} - i \frac{\Gamma_N \Gamma_S}{4 \pi} \Phi^*_{\uparrow} .
\label{pairing_amplitude1}
\end{eqnarray}
where $\Phi_{\uparrow}$ is given by Eq.~(\ref{A.277}).
%
%
In Appendix~\ref{sec:Laplace} we show, that for $\mu_N=0$ the real part of $\Phi_{\uparrow}$ vanishes.
Let us next analyze Eq.~(\ref{pairing_amplitude1}) for different initial conditions and values of the QD energy levels.
For $n_{\sigma}(0)=0$, $n_{-\sigma}(0)=1$ and $\mu_N=0$ the function $\left< \hat{d}_{\downarrow}(t)
\hat{d}_{\uparrow}(t)\right>$ is real and  non-oscillating in time and is equal to $-\frac{\Gamma_N \Gamma_S}{4 \pi}
\mbox{\rm Im} \Phi_{\uparrow}$, regardless of $\varepsilon_{\sigma}$. However,  for $\mu_N \neq 0$ also imaginary
part of $\left< \hat{d}_{\downarrow}(t) \hat{d}_{\uparrow}(t)\right>$ equals $-\frac{\Gamma_N \Gamma_S}{4 \pi}
\mbox{\rm Re} \Phi_{\uparrow}$ and is non-oscillating function. For the initial conditions $(n_{\sigma}(0),
n_{-\sigma}(0)=0)=(0,0)$ or $(1,1)$ the picture is completely different.  Depending on the value of
$\varepsilon_{\uparrow}+\varepsilon_{\downarrow}$ the real part of $\left< \hat{d}_{\downarrow}(t)
\hat{d}_{\uparrow}(t)\right>$ oscillates for  $\varepsilon_{\uparrow}+\varepsilon_{\downarrow}=0$ or is a smooth
function of time for $\varepsilon_{\uparrow}+\varepsilon_{\downarrow}\neq 0$. Simultaneously, the imaginary part of the
QD on-dot pairing oscillates irrespective of $\varepsilon_{\sigma}$. The oscillatory part of  $\left<
\hat{d}_{\downarrow}(t) \hat{d}_{\uparrow}(t)\right>$ are dumped via $e^{-\Gamma_{N}t}$ factor, emphasizing the crucial
role of continuum states of the normal electrode in relaxation processes.

\begin{figure}
\includegraphics[width=0.95\columnwidth]{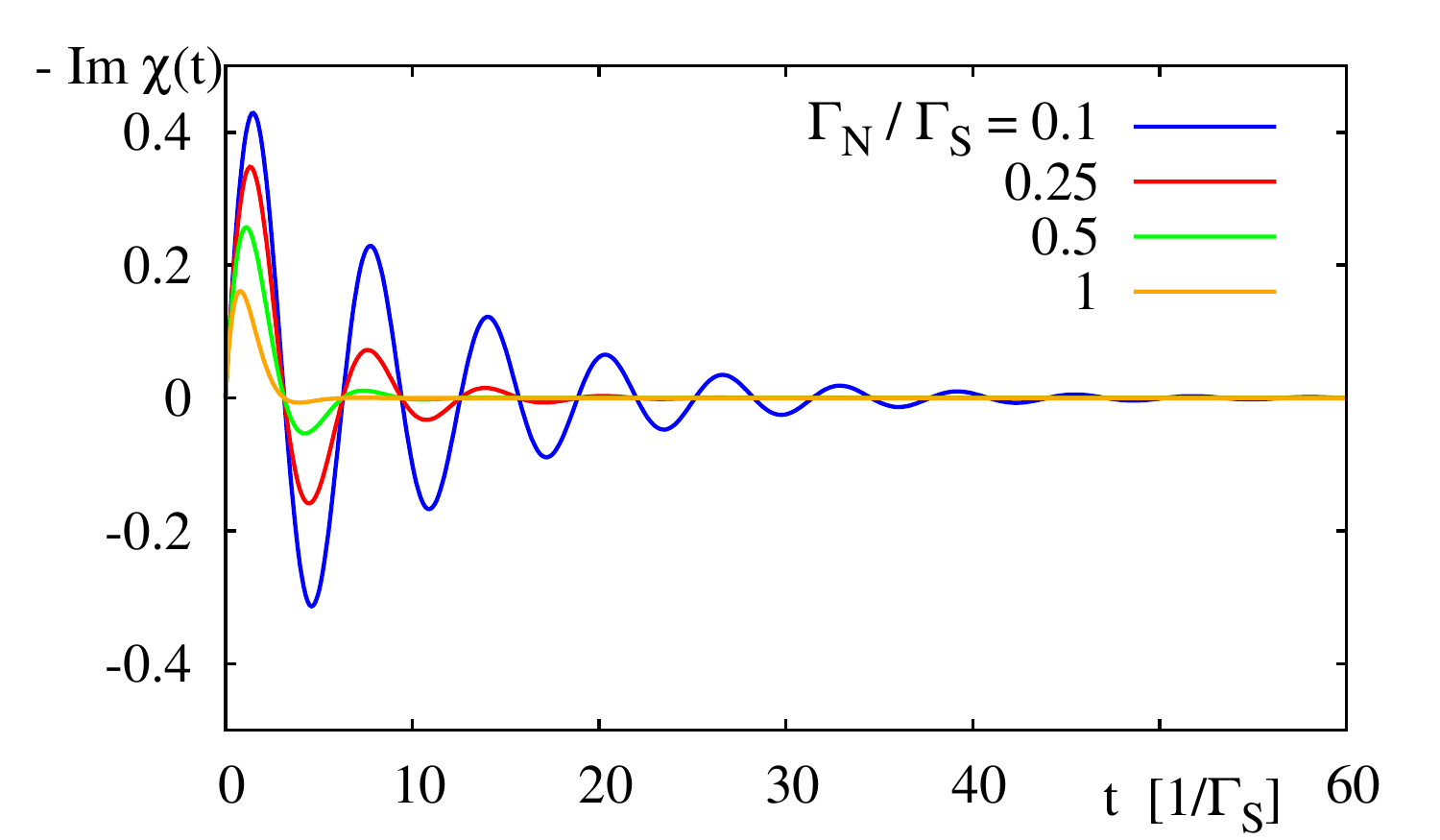}
\caption{
The imaginary part of the induced on-dot pairing
$\left< \hat{d}_{\downarrow}(t)\hat{d}_{\uparrow}(t)\right>$ obtained
for the same parameters as in Fig.\ \ref{figure2}.}
\label{figure4}
\end{figure}

\begin{figure}[t]
\includegraphics[width=0.95\columnwidth]{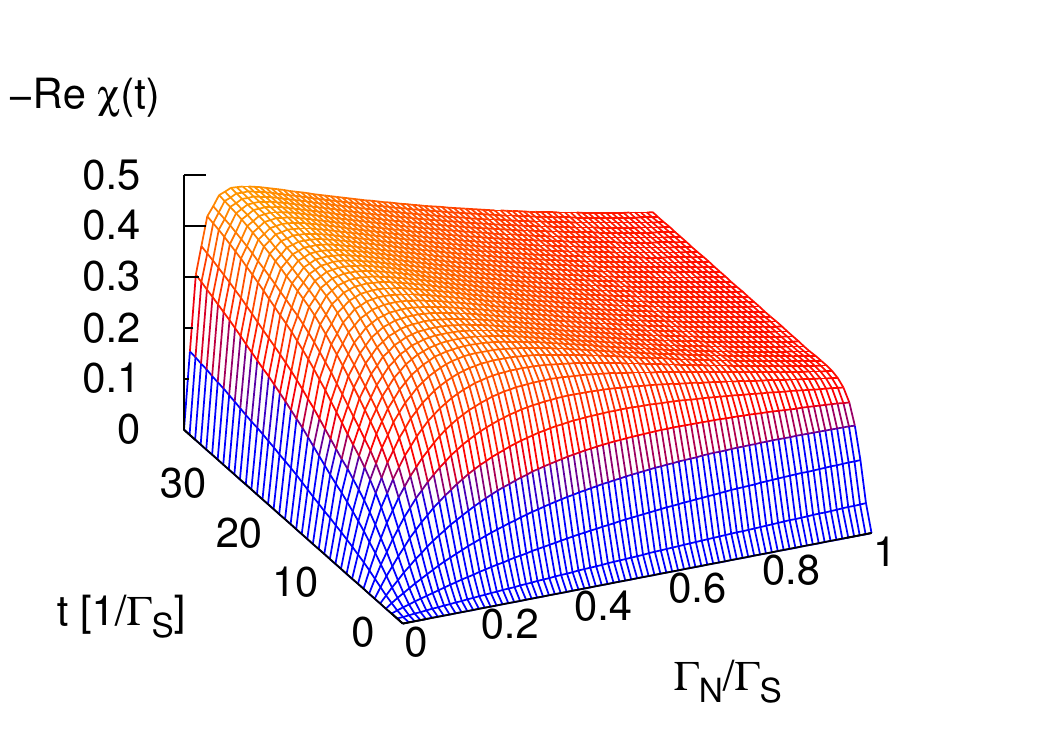}
\caption{The time-dependent real part of $\left< \hat{d}_{\downarrow}(t)
\hat{d}_{\uparrow}(t)\right>$ obtained for $\Gamma_{S}=1$ and
the same parameters as in Fig.~\ref{figure2}.}
\label{figure4b}
\end{figure}

In Fig.\ \ref{figure4} we show the imaginary part of the on-dot pairing
$\left< \hat{d}_{\downarrow}(t)\hat{d}_{\uparrow}(t)\right>$
assuming the initial QD occupancy $n_{\sigma}(0)=0$.
Period $T$ of the damped quantum oscillations depends on the excitation energy between the subgap Andreev
quasiparticles \cite{Baumgartner-17} via $T=2\pi/ \sqrt{\left(
\varepsilon_{\downarrow}+\varepsilon_{\uparrow}\right)^{2} +\Gamma_{S}^{2}}$. For $\mu_{N}=0$ these oscillations are
related to the transient current $j_{S\sigma}(t)$ flowing between the proximitized QD and the superconducting lead (see
Sec. \ref{sec:N-QD-S_current}) in analogy to the Josephson junction comprising two superconducting pieces, differing in
phase of the order parameter. On the other hand, the real part (Fig.\ \ref{figure4b}) evolves monotonously to its
asymptotic value, except of one particular case $\Gamma_{N}=0$, when the real part of $\left<
\hat{d}_{\downarrow}(t)\hat{d}_{\uparrow}(t)\right>$  vanishes.

\subsection{Transient currents for unbiased system \label{sec:N-QD-S_current}}

So far we have discussed the quantities which are important, but unfortunately they are not directly accessible
experimentally. Let us now consider the measurable currents $j_{N\sigma}(t)$ and $j_{S\sigma}(t)$, flowing from the QD
to the external leads. Formally the transient current is defined by $j_{\beta \sigma}(t)= \left<
\frac{d\hat{N}_{\beta}(t)}{dt} \right>$, where $\hat{N}_{\beta}(t)$ counts the total number of electrons in electrode
$\beta=N,S$. For instance $j_{N\sigma}(t)$ simplifies to the standard formula \cite{Jauho-1994}
\begin{eqnarray}
j_{N\sigma}(t) = 2  \; \mbox{\rm Im} \sum_{\bf k} V_{\bf k}\left<
\hat{d}^{\dagger}_{\sigma}(t) \hat{c}_{{\bf k}\sigma}(t)\right> .
\label{eqn12}
\end{eqnarray}
Assuming the energies of itinerant electrons to be static
$\varepsilon_{{\bf k}\sigma}(t)=\varepsilon_{{\bf k}\sigma}$ one obtains
\begin{eqnarray}
\hat{c}_{{\bf k}\sigma}(t) \!=\! \hat{c}_{{\bf k}\sigma}(0) e^{-i \varepsilon_{{\bf k}\sigma} t} \!\!- \! i \!\!
\int_{0}^{t}\!\! dt' V_{{\bf k}} e^{-i \varepsilon_{{\bf k}\sigma} (t-t')} \hat{d}_{\sigma}\!(t') ,
\end{eqnarray}
and within the wide-band-limit approximation it yields
\begin{eqnarray}
j_{N\sigma}(t) & = & 2 \mbox{\rm Im} \left( \sum_{\bf k} V_{\bf k}e^{-i\varepsilon_{\bf k}t} \left<
\hat{d}^{\dagger}_{\sigma}(t) \hat{c}_{{\bf k}\sigma}(0) \right> \right) - \Gamma_{N} n_{\sigma}(t) .\nonumber\\
\label{current_normal}
\end{eqnarray}
Finally, inserting the time-dependent operator $\hat{d}^{\dagger}_{\sigma}(t)$ 
[Eq.~(\ref{A.8}) to Eq.\ (\ref{eqn12})] we obtain 
\begin{eqnarray}
j_{N\sigma}(t) & = & - \Gamma_{N}n_{\sigma}(t) + \frac{\Gamma_{N}}{\pi} \;
\mbox{\rm Re} \left( \int_{-\infty}^{\infty} d\omega  f_N(\omega)
e^{-i\omega t} \right. \nonumber \\ &\times & \left. {\cal{L}}^{-1} \left\{
\frac{s+i\varepsilon_{-{\sigma}}+\Gamma_{N}/2} {(s-s_1)(s-s_2)(s-i\omega)}
\right\} (t) \right) . \label{current_N}
\end{eqnarray}
To compute the transient current of opposite spin electrons, $j_{N-\sigma}(t)$, one should replace the set of auxiliary
parameters $(s_1, s_2, s_3, s_4)$ by the following one $(s_3, s_4, s_1, s_2)$. In particular, for
$\varepsilon_{\sigma}=0$ we get
\begin{eqnarray}
j_{N\sigma}(t) &=&  \frac{\Gamma_{N}}{\pi} \; \int_{-\infty}^{\infty} d\omega f_N(\omega)
\left\{ e^{-\Gamma_{N}t/2}  \right. \label{normal_current} \\
& \times & \frac{1}{2}\sum_{p=\pm} \frac{\omega_{p}\mbox{\rm sin}\left( \omega_{p}t\right)
-\frac{\Gamma_{N}}{2}\mbox{\rm cos}\left( \omega_{p}t\right)}{\left( \frac{\Gamma_{N}}{2}\right)^{2} +\omega_{p}^{2}}
\nonumber
 \\ & + & \left.
\frac{\Gamma_{N} \left[ \left( \frac{\Gamma_{N}}{2}\right)^{2} + \left( \frac{\Gamma_{S}}{2}\right)^{2} + \omega^{2}
\right]} {\left[ \left( \frac{\Gamma_{N}}{2}\right)^{2} + \omega_{-}^{2} \right] \left[ \left(
\frac{\Gamma_{N}}{2}\right)^{2} + \omega_{+}^{2} \right] } \right\} - \Gamma_{N}n_{\sigma}(t) ,\nonumber
\end{eqnarray}
where $\omega_{\pm}=\frac{\Gamma_{S}}{2}\pm\omega$. In absence of the superconducting lead this formula is identical
with the result obtained by means of the nonequilibrium Green's function method.

\begin{figure}[t]
\includegraphics[width=0.95\columnwidth]{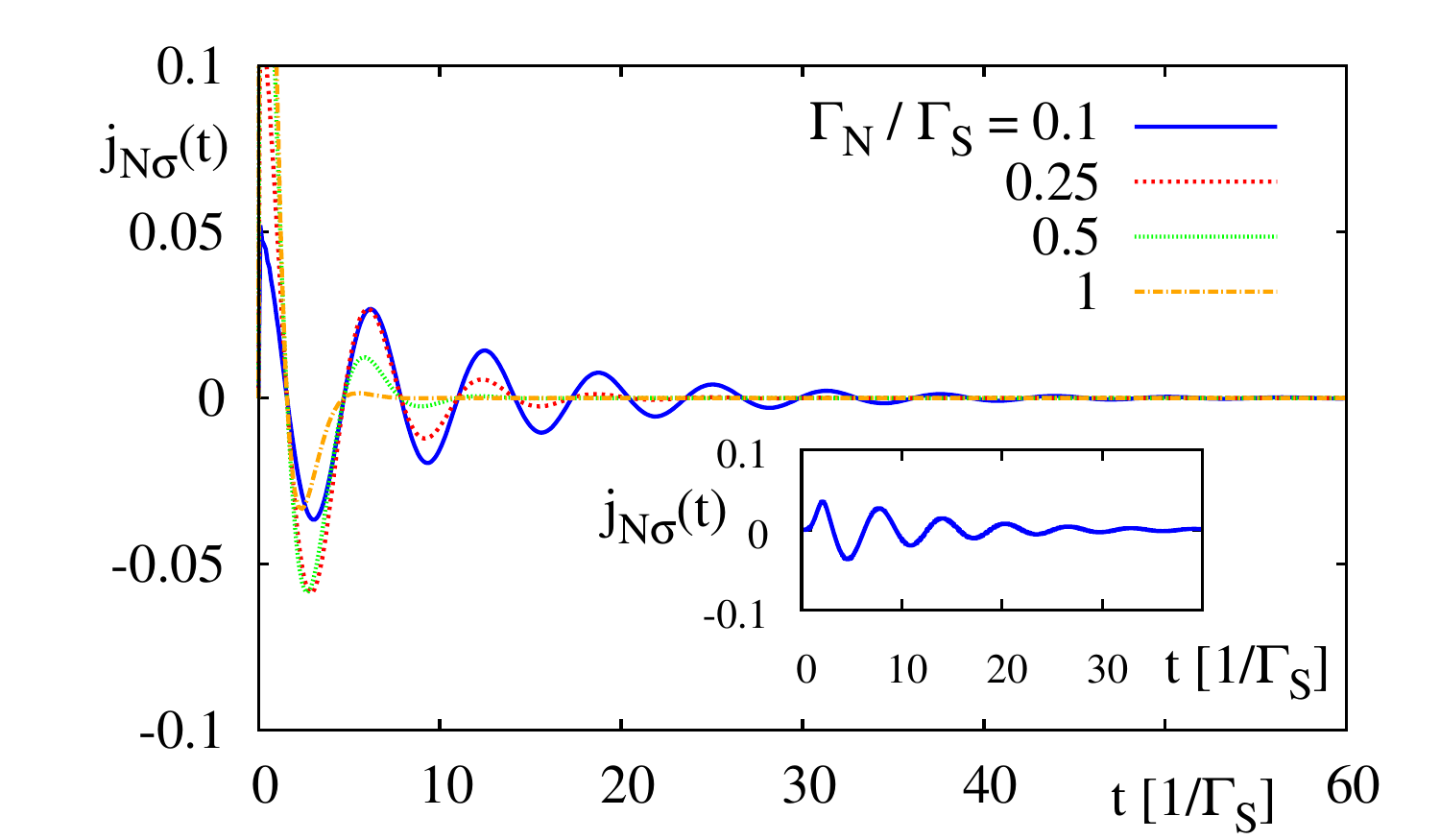}
\caption{Transient current between the QD and the normal lead induced by a sudden  coupling  in absence 
of any bias. Results are obtained for the same  parameters as in Fig.\ \ref{figure2}. The inset shows 
the transient current obtained for the sinusoidal coupling profiles $V_{{\bf k},{\bf q}}(t)$, assuming 
$\Gamma_{N}/\Gamma_{S}=0.1$.} \label{figure3a}
\end{figure}

In Fig.\ \ref{figure3a} we present transient behavior of the current $j_{N\uparrow}(t)$ induced by an abrupt coupling
of the QD to both external leads for $\mu_{N}=\mu_{S}=0$ (i.e. without any bias). Similarly to the time-dependent QD
occupancy (Fig.\ \ref{figure2}) we observe the quantum oscillations of  the period $2\pi/\Gamma_{S}$ exponentially
decaying with the envelope coefficient $e^{-\Gamma_{N}t}$. Large value of the current at $t=0^{+}$ is a
consequence of the abrupt switching (\ref{abrupt_coupling}). One may ask whether this instantenous switching
could be realistic in experimental situations. To check if any smooth (gradual) coupling process would 
affect our main conclusions we have computed the transient currents, assuming the sinusoidal switching 
profile $V_{{\bf k},{\bf q}}(t)=\frac{V_{{\bf k},{\bf q}}}{2} \left( \sin{(\pi \Gamma_{N} t - \pi/2)}+1 \right)$
for $0<t\leq 1/\Gamma_{N}$ and keeping constant value $V_{{\bf k},{\bf q}}$ for $t>1/\Gamma_{N}$.
We have solved this problem numerically and present some representative results (for $\Gamma_{N}/\Gamma_{S}=0.1$) in 
the inset in Fig.\ \ref{figure3a}. We noticed, that for $t>1/\Gamma_{N}$  all the time-dependent 
quantities are not particularly affected. The only difference (in comparison to the abrut coupling) 
is in the early time region $0<t<1/\Gamma_{N}$. For instance, the transient current $j_{N\uparrow}(t)$
smoothely evolves from zero to its asymptotic behavior with the same period of quantum oscillations.

In similar steps, we have also determined the transient  current $j_{S\sigma}(t)=2  \; \mbox{\rm Im} \sum_{\bf k}
V_{\bf q}\left< \hat{d}^{\dagger}_{\sigma}(t) \hat{c}_{{\bf q}\sigma}(t)\right>$. Effective quasiparticles in
superconductors are represented by a coherent superposition of the particle and hole degrees of freedom, so for this
reason the time-dependent operator $\hat{c}_{{\bf q}\sigma}(t)$ consists of four contributions (see Eq.\ \ref{A.10}).
Final expression for $j_{S\sigma}(t)$ becomes rather lengthy, therefore we present it  in
Appendix \ref{sec:LaplaceA4}. However, in absence of external voltage the current (\ref{current_j_S}) simplifies to
\begin{eqnarray}
j_{S\sigma}(t) = \frac{{\Gamma}^{2}_S}{2\sqrt{\delta}} \mbox{\rm sin} \left( \sqrt{\delta}t\right) e^{-\Gamma_{N}t}
\left[ 1 \!-\! \sum_{\sigma'} n_{\sigma'}(0)  \right] . \label{simple_j_S}
\end{eqnarray}
When  the QD is initially empty/full the transient current $j_{S\sigma}= \pm
\frac{{\Gamma}^{2}_S}{2\sqrt{\delta}}\mbox{\rm sin} \left( \sqrt{\delta}t\right)
e^{-\Gamma_{N}t}$ reveals the damped oscillations. Contrary to this behavior, for
the initial occupancies  $n_{\sigma}(0)=0$ and $n_{-{\sigma}}(0)=1$ the current
(\ref{simple_j_S})  vanishes. We assign this feature to inefficiency of the proximity
effect whenever the QD is singly occupied, because electron pairing operates only
by mixing the empty with the doubly occupied QD configurations. Initial
conditions have thus important influence on transient phenomena.

Furthermore, Eq.~\ref{A.23} for $\left< \hat{d}_{\downarrow}(t)\hat{d}_{\uparrow}(t)\right>$ and Eq.~\ref{current_j_S}
imply the  exact relationship $j_{S\sigma}(t) = -\Gamma_{S} \mbox{\rm Im} \left<
\hat{d}_{\downarrow}\hat{d}_{\uparrow} \right>$ which is popular in considerations of charge transport through
Josephson junctions \cite{Zonda-2015}. The transient current $j_{S\sigma}(t)$ can hence be simply inferred from Fig.\
\ref{figure4}. At this level it is important to remark, that the charge conservation  of our system is correctly
satisfied, i.e.
\begin{eqnarray}
j_{S\sigma}(t) + j_{N\sigma}(t)=\frac{d}{dt} n_{\sigma}(t) .
\end{eqnarray}
%

\subsection{Transient currents for biased system \label{sec:N-QD-S_current2}}

We have seen so far, that time-dependent QD occupancy and transient currents provide indirect information about the
subgap quasiparticle energies and dynamical transitions between them. In absence of any voltage ($\mu_{N}=\mu_{S}=0$)
these transient currents finally vanish, with a rate dependent on the relaxation processes caused by the coupling
$\Gamma_{N}$ with a continuum of metallic lead. From the practical point of view, much more convenient way for probing
the time-scales characteristic for the Andreev/Shiba quasiparticles could be provided by transient properties of the
biased system $\mu_{N} \neq \mu_{S}$. Following the steps discussed in previous Sec.\ \ref{sec:N-QD-S_current} we shall
study here the time-dependent differential conductance $G_{\sigma}(\mu,t)\equiv \frac{d}{d\mu} j_{N\sigma}(t)$ as a
function of external voltage $\mu\equiv\mu_{N}$ (throughout this work the superconducting lead is assumed to be
grounded $\mu_{S}\!=\!0$). At zero temperature Eq.\ (\ref{current_N}) implies
\begin{widetext}
\begin{eqnarray}
G_{\sigma}(\mu,t) & = & \Gamma_{N} \; \mbox{\rm Re} \left[ e^{-i\mu t} {\cal{L}}^{-1} \left\{
\frac{s+i\varepsilon_{-\sigma} +\Gamma_{N}/2}{(s-s_{1})(s-s_2)(s-i\mu)} \right\} \!(t)
\right] \label{conductance} \\
& - & \frac{\Gamma_{N}^{2}}{2} \; {\cal{L}}^{-1} \! \left\{ \frac{s+i\varepsilon_{-\sigma}
+\Gamma_{N}/2}{(s-s_{1})(s-s_2)(s-i\mu)} \right\} \!(t) \;\; {\cal{L}}^{-1} \! \left\{ \frac{s-i\varepsilon_{-\sigma}
+\Gamma_{N}/2}{(s-s_{3})(s-s_4)(s+i\mu)} \right\}\!(t)
\nonumber \\
& + &  \frac{\Gamma_N^2\Gamma_{S}^2}{8} \; {\cal{L}}^{-1}\!\left\{ \frac{1}{(s-s_{1})
(s-s_2)(s+i\mu)}\right\}\!(t) \;\; {\cal{L}}^{-1} \! \left\{ \frac{1}{(s-s_{3})(s-s_4)
(s-i\mu)} \right\} \!(t) , \nonumber
\end{eqnarray}
\end{widetext}

\begin{figure}
\includegraphics[width=0.9\columnwidth]{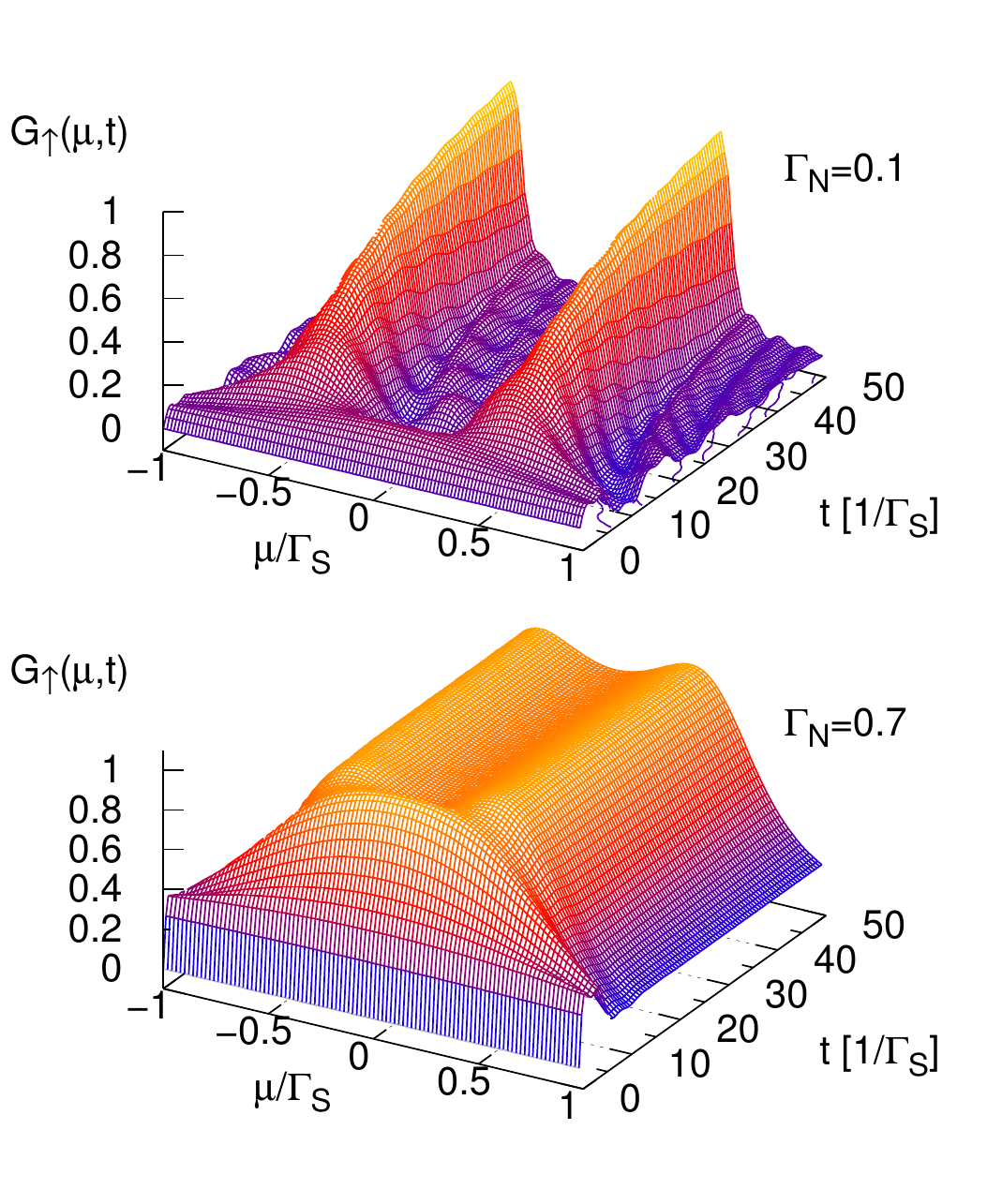}
\caption{The time-dependent differential conductance $G_{\uparrow}(\mu,t)=G_{\downarrow}(\mu,t)$ 
(in units of $\frac{4e^2}{h}$) obtained for $\varepsilon_{\sigma}=0$, 
$\Gamma_{N}=0.2$ (top panel) and $\Gamma_{N}=0.7$ (bottom panel).}
\label{cond_plot}
\end{figure}

\noindent 
where the conductance is expressed in units of $\frac{2e^2}{h}$. Expression for $G_{\downarrow}(\mu,t)$
can be obtained by the replacement $(s_{1},s_{2},s_{3},s_{4}) \rightarrow (s_{3},s_{4},s_{1},s_{2})$. 
Using the corresponding inverse Laplace transforms we find (for $\varepsilon_{\sigma}=0$,
$G_{\uparrow}=G_{\downarrow}=G$):
\begin{eqnarray}
G(\mu,t) &=& \Gamma_{N} \left\{ { e^{-\Gamma_N t/2} \over 2} \sum_{p=+,-} \frac{\mu_p \sin(\mu_p t)- {\Gamma_N
\over 2}\cos(\mu_p t)} {\left(\frac{\Gamma_{N}}{2}\right)^{2}+\mu_p^2} \right. \nonumber\\
&+& \left. \frac{\left(\frac{\Gamma_{N}}{2}\right)
\left[\left(\frac{\Gamma_{N}}{2}\right)^2+\left(\frac{\Gamma_{S}}{2}\right)^2+\mu^2  \right] } {\left[
\left(\frac{\Gamma_{N}}{2}\right)^{2} +\mu_+^2 \right] \left[ \left(\frac{\Gamma_{N}}{2}\right)^{2} +\mu_-^2 \right] }
\right\}
\label{conductance2} \\
&-& \frac{\Gamma_{N}^2}{2} F_1(\mu,t) + \frac{\Gamma_{N}^2\Gamma_S^2}{8} F_2(\mu,t) , \nonumber
\end{eqnarray}
where $F_1(\mu,t)$ and $F_2(\mu,t)$ are given in Eqs.~(\ref{A.18},\ref{A.19}), and
$\mu_{+/-}=\mu\pm \Gamma_S/2$. In the steady limit, $t\rightarrow \infty$ and for $\varepsilon_{\sigma}=0$, keeping
only terms that survive at late times, we obtain the expression identical with the result derived for the same setup
within the B\"uttiker-Landauer approach \cite{Domanski-2008}
\begin{eqnarray}
G(\mu,\infty) =
\frac{\Gamma_{N}^{2}  \Gamma_{S}^{2}} {4\left[ \left(\frac{\Gamma_{N}}{2}\right)^{2} +\mu_-^{2}\right] \left[
\left(\frac{\Gamma_{N}}{2}\right)^{2} +\mu_+^{2}\right]} .\label{conductance3}
\end{eqnarray}
For $\Gamma_S \gg \Gamma_N$ the local extrema of this expression occur at $\mu=\pm \frac{\Gamma_{S}}{2}$ and they
correspond to the energies of subgap bound states. For an arbitrary set of model parameters such information is encoded
in Eq.\ (\ref{conductance}) which quantitatively specifies development of the in-gap states driven by the sudden
switching at $t=0$. In Fig.\ \ref{cond_plot} we present the differential conductance obtained numerically for
$\Gamma_{N}/ \Gamma_{S}=0.1$ and $0.7$. Let us notice, that differential conductance approaches its steady-limit shape
$G_{\uparrow}(\mu,t=\infty)$ characterized by two Lorentzian quasiparticle peaks centered at $\sim \pm
\frac{\Gamma_S}{2}$. Their broadening $\Gamma_{N}$ is related to the inverse life-time.

\begin{figure}[b]
\includegraphics[width=0.9\columnwidth]{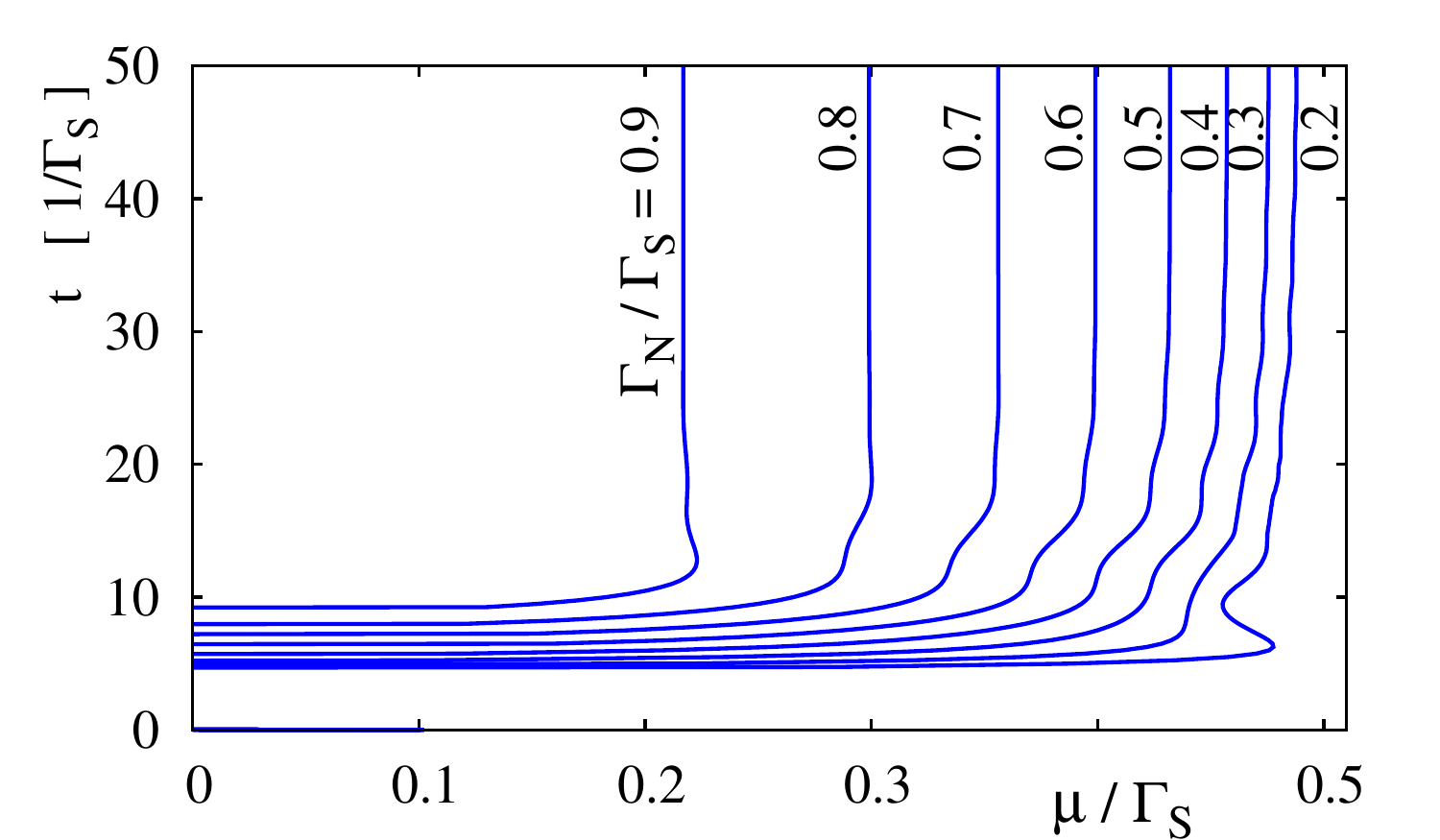}
\caption{Positions of the quasiparticle maxima vs. time and $\mu/\Gamma_S$ appearing in the differential conductance
$G_{\uparrow}(\mu,t)$ for a number of ratios $\Gamma_{N}/\Gamma_{S}$, as indicated. For negative values of
$\mu/\Gamma_S$ the results are symmetrical.} \label{profiles}
\end{figure}

More careful examination of $G_{\uparrow}(\mu,t)$ indicates, that development of the subgap quasiparticles
proceeds in three steps with typical time-scales $\tau_{1}$,  $\tau_{2}$ and $\tau_{3}$, as can be deduced from Fig.\
\ref{profiles} and \ref{cond_plot}.  in Fig.~\ref{profiles} we show how the position of the quasiparticle maxima
develops in time for different $\Gamma_N$.  At $t\!=\!\tau_{1}$ there emerge two maxima from the single broad structure
where $\tau_1$ changes approximately from 5 (for $\Gamma_N=0.2$) up to 10 (for $\Gamma_N=0.9$) units of time. These
maxima move rapidly (essentially during 1-2 units) from $\mu=0$ up to some value of $\mu$ which depends on $\Gamma_N$.
Next, the position of the quasiparticle peaks evolve continuously to their steady limit position
$\mu=\pm\sqrt{\Gamma_S^2-\Gamma_N^2}$ with $\tau_2$ approximately changing from 15 (for $\Gamma_N=0.9$) up to 30 (for
$\Gamma_N=0.2$) units of time. Finally, the asymptotic quasiparticle feature is achived with the evolve function $1-
\mbox{\rm exp}\left(-t/\tau_{f} \right)$ where $\tau_f=2/\Gamma_N$, see Eqs.~(\ref{conductance2},\ref{A.18},\ref{A.19})
where the terms proportional to ${\rm exp}\left(-\Gamma_N t/2 \right)$ are responsible for such asymptotic behaviour.
We also clearly see, that near the quasiparticle peaks the total differential conductance $\sum_{\sigma}G(\mu,t
\rightarrow \infty)$ acquires its optimal value $4e^{2}/h$ known from the previous studies (see e.g. the Ref.\
\onlinecite{Rodero-11}).

\section{Correlation effects\label{sec:Correlations}}

Local repulsive interactions $U\hat{n}_{\uparrow}\hat{n}_{\downarrow}$
compete with the proximity-induced electron pairing. This issue has been addressed
in the steady limit by numerous methods \cite{Rodero-11}. In particular, it has
been shown \cite{Bauer-2007} that effective pairing (manifested by the in-gap
states) is predominantly sensitive to the ratio $U/\Gamma_{S}$ and depends
on the energy level $\varepsilon_{\sigma}$. Various experimental realizations
of the correlated quantum dot in N-QD-S geometry \cite{Deacon-10,Lee-12,Pillet-13,Zitko-2015}
indicated that the Coulomb potential $U$ safely exceeds (at least one order of magnitude)
the superconducting energy gap $\Delta$. Under such circumstances the correlation
effects show up in the subgap regime $|\omega|<\Delta$ merely by a quantum
phase transition (or crossover) from the spinless (BCS-type) state $u\left| 0 \right>
+ v \left| \uparrow \downarrow \right>$ to the spinful (singly occupied) configuration
$\left| \sigma \right>$. This changeover occurs upon increasing the ratio $U/\Gamma_{S}$
and  above some critical value of the Coulomb potential $U_{cr}$ there
can be observed  the subgap Kondo effect (even in the superconducting atomic limit)
\cite{Zitko-2015,Domanski-2016}. We shall briefly analyze some correlation effects,
focusing on the transient effects.

\subsection{Competition between pairing and correlations\label{sec:Competition}}

The aforementioned quantum phase transition can be qualitatively captured already
within the lowest order (Hartree-Fock-Bogoliubov) decoupling scheme
\begin{eqnarray}
\hat{d}^{\dagger}_{\uparrow}\hat{d}_{\uparrow}
\hat{d}^{\dagger}_{\downarrow}\hat{d}_{\downarrow} & \simeq &
n_{\uparrow}(t) \hat{d}^{\dagger}_{\downarrow}\hat{d}_{\downarrow}
+n_{\downarrow}(t) \hat{d}^{\dagger}_{\uparrow}\hat{d}_{\uparrow}
-n_{\uparrow}(t)n_{\downarrow}(t)
\nonumber \\ & + &
\chi^{*}(t) \hat{d}^{\dagger}_{\uparrow} \hat{d}^{\dagger}_{\downarrow}
+\chi(t) \hat{d}_{\downarrow} \hat{d}_{\uparrow}
-\left| \chi(t) \right|^{2} .
\label{HFB}
\end{eqnarray}
Using this approximation (\ref{HFB}) one can incorporate the Hartree-Fock terms  into the renormalized energy level
$\tilde{\epsilon}_{\sigma} \equiv \epsilon_{\sigma}+U n_{-\sigma}(t)$, whereas the anomalous (pair source and drain)
terms rescale the effective pairing potential $\tilde{\Gamma}_{S}/2 \equiv \Gamma_{S}/2 + U \chi(t)$. This decoupling
procedure (\ref{HFB}) can give a crossing of the subgap quasiparticle energies at some critical ratio $U/\Gamma_{S}$,
dependent also on $\varepsilon_{\sigma}$. In the Josephson junctions such effect would cause a reversal of the d.c.\
tunneling current, so called, $0-\pi$ transition \cite{Zonda-2015,Zonda-2016}. In our N-QD-S heterostructure its
influence is noticeable, but rather less spectacular.

\begin{figure}[t]
\includegraphics[width=0.9\columnwidth]{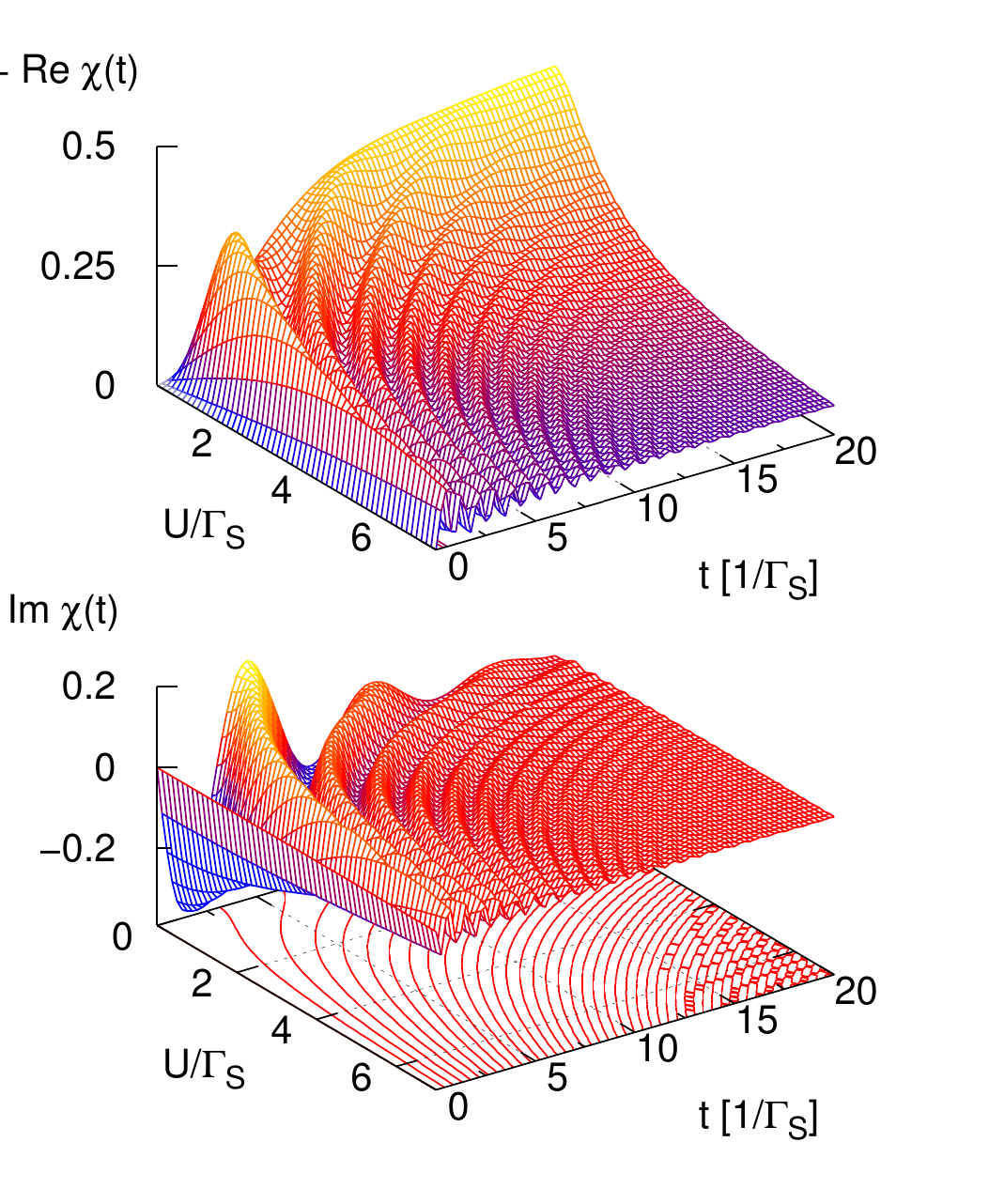}
\caption{Influence of the Coulomb potential $U$ on the real (upper panel)
and imaginary (bottom panel) parts of the induced pairing $\chi(t)=
\langle \hat{d}_{\downarrow}\hat{d}_{\uparrow}\rangle$ obtained for
the unbiased system, using $\varepsilon_{\sigma}=0$, $\Gamma_{N}=0.2$,
and $\Gamma_{S}\equiv 1$.} \label{corr_pairing}
\end{figure}

Analytical determination of the dynamical observables (discussed in Sec.~\ref{sec:N-QD-S}) is unfortunately not
feasible in the present case, because the renormalized energy level $\tilde{\varepsilon}_{\sigma}(t)$ and effective
pairing potential $\tilde{\Gamma}_{S}(t)$ are time-dependent in nonexplicit way and the method used in the previous
section is useful only for consideration of systems with constant QD energy levels and couplings with the leads.
Therefore, in what follows, we consider the Coulomb repulsion in the system of the  proximitized QD
coupled only to the normal lead, applying the Hartree-Fock-Bogoliubov
approximation (\ref{HFB}). We have computed numerically
$n_{\sigma}(t)$, $\langle \hat{d}_{\downarrow}(t) \hat{d}_{\uparrow}(t)\rangle$ and $j_{N\sigma}(t)$, solving the
closed set of differential equations for time-dependent functions  $n_{\sigma}(t)$ and $\langle \hat{d}_{\downarrow}(t)
\hat{d}_{\uparrow}(t)\rangle$, respectively (see Appendix B).
At intermediate steps we had to compute  additionally the  expectation values $\langle \hat{d}^{\dagger}_{\sigma}(t)
c_{{\bf k} \sigma}(0)\rangle$ and $\langle \hat{d}_{\sigma}(t) \hat{c}_{{\bf k} -\sigma}(0)\rangle$. All these
quantities have been determined within the Runge-Kutta numerical algorithm.

\begin{figure}
\includegraphics[width=0.8\columnwidth]{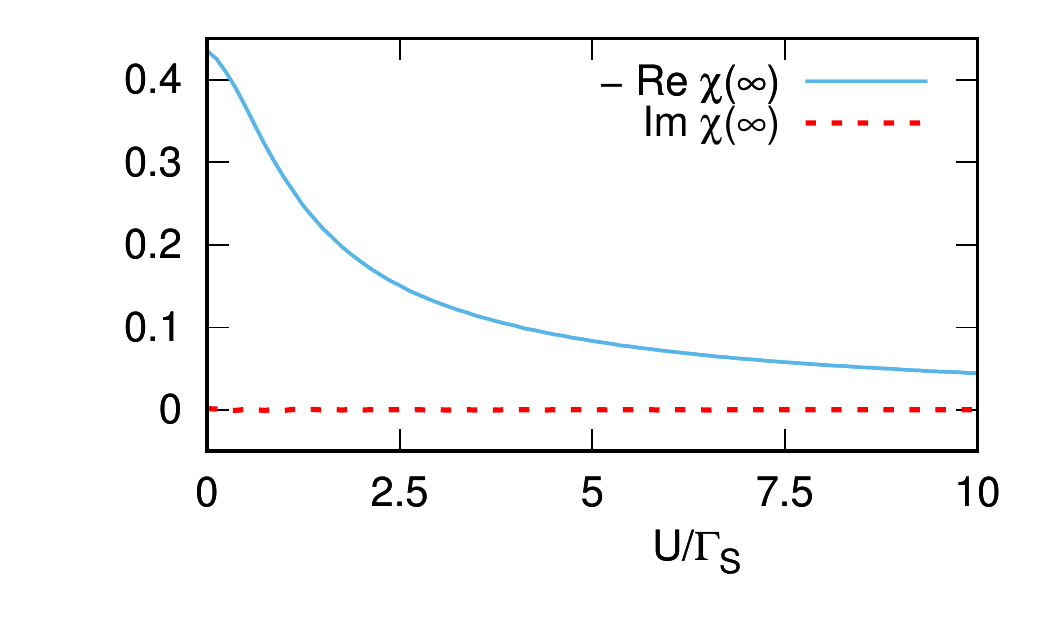}
\caption{Asymptotic value of complex on-dot pairing $\chi(t\rightarrow\infty)$ suppressed by the Coulomb repulsion $U$
obtained for the same model parameters as in Fig.~\ref{corr_pairing}.}
\label{corr_competition}
\end{figure}

Fig.\ \ref{corr_pairing} displays influence of the Coulomb potential $U$ on the induced order parameter $\chi(t)$  for
the unbiased system. The imaginary part, which is strictly related to the transient current, exhibits the damped
quantum oscillations. Their period and amplitude are substantially suppressed by the Coulomb potential. We assign this
fact to a competition between the on-dot pairing and local Coulomb repulsion. The real part of $\chi(t)$ is
characterized by the same quantum oscillations. The asymptotic  value of the complex order parameter
$\chi(t\rightarrow\infty)$ with respect to the Coulomb potential $U$ is shown in Fig.\ \ref{corr_competition} for
$\varepsilon_{\sigma}=0$, $\Gamma_{N}/\Gamma_{S}=0.2$. Such monotoneously decreasing Re$\chi(\infty)$ confirms a
competing relationship between the on-dot pairing and the local repulsion.

\begin{figure}
\includegraphics[width=0.9\columnwidth]{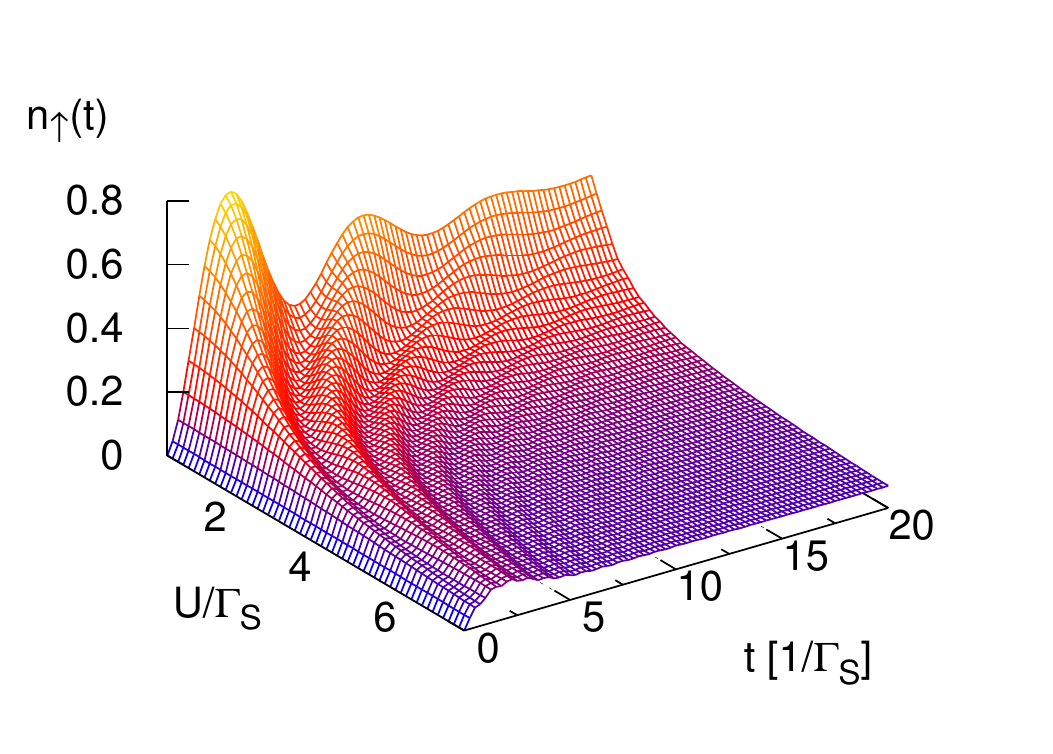}
\caption{The time-dependent occupancy of the correlated quantum dot
 for $\epsilon_{\sigma}=0$, $\Gamma_{N}=0.2$, $\Gamma_{S}=1$ in
absence of external voltage.}
\label{corr_filling}
\end{figure}

In Fig.\ \ref{corr_filling} we show influence of the Coulomb potential $U$ on
the QD occupancy $n_{\uparrow}(t)$. Besides the quantum oscillations, similar
to the ones observed in the complex order parameter (Fig.\ \ref{corr_pairing}),
we notice a partial reduction of the QD charge upon increasing $U$. Apparently
this is caused by the Hartree term $U n_{-\sigma}(t)$, that lifts the renormalized
QD level $\tilde{\varepsilon}_{\sigma}(t)$.
In Appendix \ref{C} we briefly  discuss the time-dependent subgap Kondo effect.

\section{Summary\label{sec:Summary}}

We have investigated transient effects driven by a sudden coupling of the quantum dot
to the metallic and superconducting leads. Our study has revealed a gradual buildup
of the subgap Andreev quasiparticle states, which is controlled by the coupling $\Gamma_{N}$
to a continuous spectrum of the metallic lead. Depending on the initial
quantum dot occupancy, we have also found the damped quantum oscillations of
the charge occupancy $n_{\sigma}(t)$, the complex order parameter $\chi(t)$ and
the transient currents $j_{N\sigma}(t)$, $j_{S\sigma}(t)$. Period of these oscillations 
would be sensitive to the Andreev quasiparticle energies which can be indirectly  controlled 
via a coupling $\Gamma_{S}$ to the superconducting reservoir.

Analogous effects (relaxation and quantum oscillations) have been recently 
reported in Refs [\onlinecite{LevyYeyati-2017}] and [\onlinecite{LevyYeyati-2016}] in
studies of the metastable subgap states for the Josephson junction, considering finite value
of the superconducting gap. We estimate that in realistic systems, where $\Gamma_{S} \sim 0.2$ meV, 
the period of quantum oscillations would be a fraction of nanoseconds or in picoseconds
regime (hence should be empirically detectable). Buildup of the subgap Andreev 
quasiparticle states is expected to be formed in N-QD-S junctions on much 
longer time-scale, corresponding to a microsecond regime. Our estimations seem 
to reliable, when comparing them with dynamical transitions between the subgap 
bound states of nanotubes \cite{Baumgartner-17} and parity switchings observed 
in the superconducting atomic contacts \cite{experiment-15}.

We also addressed the correlation effects by means of the Hartree-Fock-Bogoliubov
approximation, revealing that the repulsive Coulomb potential $U$  suppresses 
the proximity-induced electron pairing. We have explored some time-dependent 
signatures of this competition. In particular, we have found that $\Gamma_{N}$ 
controls the rate at which the stationary limit behavior is achieved, whereas 
period of the damped quantum oscillations is dependent on the Coulomb potential 
due to its influence of the Andreev quasiparticle energies \cite{Bauer-2007}. 

Finally, we have tried to evaluate the characteristic time-scale $\tau_{K}$ 
needed for the subgap Kondo effect to develop. Upon approaching the quantum 
phase transition from the (spinful) doublet side we predict the strong reduction 
of this scale $\tau_{K}$, originating from a subtle interplay between the 
induced on-dot pairing and the Coulomb repulsion \cite{Zitko-2015,Domanski-2016}. 
We hope that such variety of dynamical effects of the proximitized quantum dots 
could be verified experimentally.

\begin{acknowledgments}
We acknowledge instructive discussions with V.\ Jani\v{s} and thank A. Baumgartner 
for useful remarks on observability of the transient effects in multi-terminal
heterostructures. We also kindly thank T.\ Kwapi\'nski for technical assistance.
This work is supported by the National Science Centre (Poland) 
through the grant DEC-2014/13/B/ST3/04451 (TD).
\end{acknowledgments}

\appendix

\section{ \label{sec:Laplace}}
\label{A}
In this Appendix we derive the Laplace transforms $\hat{d}_{\sigma}(s)$ and $\hat{c}_{{\bf q}
\sigma}(s)$ which are needed for calculating the statistically averaged physical quantities 
considered in this work. We present the explict formulas for the QD occupancy, the pair 
correlation function and the transient currents flowing between QD and external reservoirs
for the case $\Delta = \infty$ and $U=0$.

\subsection{Laplace transforms\label{sec:LaplaceA1}} \label{A1}

To calculate the expectation values of the quantities studied in this work we need the time-dependent operator
$\hat{d}_{\sigma}(t)$. We can find it taking the corresponding inverse Laplace transform: ${\cal{L}}^{-1} \! \left\{
\hat{d}_{\sigma}(s) \right\}(t)$. To find $\hat{d}_{\sigma}(s)$ we write, in the first step, the equations of motion
for the closed set of eight operators: $\hat{d}_{\sigma}(t)$, $\hat{d}_{-\sigma}^{\dagger}(t)$, $\hat{c}_{{\bf k}
\sigma}(t)$, $\hat{c}_{{\bf k} -\sigma}^{\dagger}(t)$, $\hat{c}_{{\bf q} \sigma}(t)$, $\hat{c}_{-{\bf q}
-\sigma}^{\dagger}(t)$, $\hat{c}_{{\bf q} -\sigma}^{\dagger}(t)$ and $\hat{c}_{-{\bf q} \sigma}(t)$. Performing the
Laplace transforms of these differential equations we have e.g. for $\sigma=\uparrow$ (for arbitrary energy gap
$\Delta$ neglecting the correlation effects, $U=0$):
\begin{subequations} \label{EOM}
\begin{eqnarray} \label{EOM.A}
(s+i\varepsilon_{\uparrow}) \hat{d}_{\uparrow}(s) &=& -i \sum_{{\bf k}/{\bf q}} V_{{\bf k}/{\bf q}} \hat{c}_{{{\bf
k}/{\bf q}}\uparrow}(s)+\hat{d}_{\uparrow}(0) , \\
\label{EOM.B} (s+i\varepsilon_{\bf k}) \hat{c}_{{\bf k}\uparrow}(s) &=& -i V_{\bf k} \hat{d}_{\uparrow}(s) +
\hat{c}_{{\bf k}\uparrow}(0) , \\
\label{EOM.C} (s+i\varepsilon_{\bf q}) \hat{c}_{{\bf q}\uparrow}(s) &=& -i V_{\bf q} \hat{d}_{\uparrow}(s) -i\Delta
\hat{c}_{-{\bf q}\downarrow}^{\dagger}(s)+\hat{c}_{{\bf q}\uparrow}(0) , \nonumber\\ \\
\label{EOM.D} (s-i\varepsilon_{\bf q}) \hat{c}_{-{\bf q}\downarrow}^{\dagger}(s) &=& i V_{\bf q}
\hat{d}_{\downarrow}^{\dagger}(s) -i\Delta \hat{c}_{{\bf q}\uparrow}(s)+\hat{c}_{-{\bf q}\downarrow}^{\dagger}(0) ,
\nonumber\\ \\
\label{EOM.E} (s-i\varepsilon_{\downarrow}) \hat{d}_{\downarrow}^{\dagger}(s) &=& i \sum_{{\bf k}/{\bf q}} V_{{\bf
k}/{\bf q}} \hat{c}_{{{\bf k}/{\bf q}}\downarrow}^{\dagger}(s)+\hat{c}^{\dagger}_{\downarrow}(0) , \\
\label{EOM.F} (s-i\varepsilon_{\bf k}) \hat{c}_{{\bf k}\downarrow}^{\dagger}(s) &=& i V_{\bf k}
\hat{d}_{\downarrow}^{\dagger}(s) + \hat{c}_{{\bf k}\downarrow}^{\dagger}(0) , \\
\label{EOM.G} (s-i\varepsilon_{\bf q}) \hat{c}_{{\bf q}\downarrow}^{\dagger}(s) &=& i V_{\bf q}
\hat{d}_{\downarrow}^{\dagger}(s) -i\Delta \hat{c}_{-{\bf q}\uparrow}(s)+\hat{c}_{{\bf q}\downarrow}^{\dagger}(0) ,
\nonumber\\ \\
\label{EOM.H} (s+i\varepsilon_{\bf q}) \hat{c}_{-{\bf q}\uparrow}(s) &=& -i V_{\bf q} \hat{d}_{\uparrow}(s) -i\Delta
\hat{c}_{{\bf q}\downarrow}^{\dagger}(s)-\hat{c}_{-{\bf q}\uparrow}(0) . \nonumber\\
\end{eqnarray}
\end{subequations}
Here we have assumed $\varepsilon_{{\bf q}\sigma}=\varepsilon_{\bf q}$, $\varepsilon_{\bf q}=\varepsilon_{-\bf q}$ and
the subscript ${\bf k} ({\bf q})$ corresponds to the normal (superconducting) electrode.

To find $\hat{d}_{\uparrow}(s)$ we calculate $\hat{c}_{-{\bf q}\downarrow}^{\dagger}(s)$ from Eq.~(\ref{EOM.D}) and
insert it to $\hat{c}_{{\bf q}\uparrow}(s)$ obtained from Eq.~(\ref{EOM.C}). In the next step we insert  $\hat{c}_{{\bf
q}\uparrow}(s)$ into the expression for $\hat{d}_{\uparrow}(s)$ taken from Eq.~\ref{EOM.A} together with $\hat{c}_{{\bf
k}\uparrow}(s)$ obtained from Eq.~(\ref{EOM.B}). In result we have:
\begin{eqnarray} \label{A.2}
\hat{d}_{\uparrow}(s) M^{(+)}_{\uparrow}(s) &=&  \hat{A}(s)-iK(s)\hat{d}_{\downarrow}^{\dagger}(s) .
\end{eqnarray}
We next repeat the procedure (\ref{EOM.E}-\ref{EOM.H}), obtaining
\begin{eqnarray} \label{A.3}
\hat{d}_{\downarrow}^{\dagger}(s) M^{(-)}_{\downarrow}(s) &=&  \hat{B}(s)-iK(s)\hat{d}_{\uparrow}(s) ,
\end{eqnarray}
where
\begin{eqnarray}
M^{(\pm)}_{\sigma}(s) & = & s\pm i\varepsilon_{\sigma} +\sum_{\bf k} \frac{V_{\bf k}^{2}}{s\pm i \varepsilon_{\bf k}}
+\sum_{\bf q} \frac{V_{\bf q}^{2}(s\mp i \varepsilon_{\bf q})} {s^{2}+ \varepsilon^{2}_{\bf q}+\Delta^{2}} ,
\nonumber\\ \\
K(s) & = & \sum_{\bf q} \frac{V_{\bf q}^{2} \; \Delta}{s^{2} + \varepsilon_{\bf q}^{2} + |\Delta|^{2}} ,
\\
\hat{A}(s) & = & -i\sum_{\bf k} \frac{V_{\bf k} \; \hat{c}_{{\bf k}\uparrow}(0)} {s+i\varepsilon_{\bf k}} - \sum_{\bf
q} \frac{V_{\bf q}}{s^{2}+\varepsilon^{2}_{\bf q}+\Delta^{2}}
\\
& \times & \left(\Delta \hat{c}^{\dagger}_{-{\bf q}\downarrow}(0) + i (s-i\varepsilon_{\bf q}) \hat{c}_{{\bf
q}\uparrow}(0)\right) + \hat{d}_{\uparrow}(0), \nonumber
\\
\hat{B}(s) & = & i\sum_{\bf k} \frac{V_{\bf k}\hat{c}^{\dagger}_{{\bf k}\downarrow}(0)} {s-i\varepsilon_{\bf k}} +
\sum_{\bf q} \frac{V_{\bf q}}{s^{2}+\varepsilon^{2}_{\bf q}+\Delta^{2}}
\\
& \times & \left( \Delta \hat{c}_{-{\bf q}\uparrow}(0) + i (s+i\varepsilon_{\bf q}) \hat{c}_{{\bf
q}\downarrow}^{\dagger}(0)\right) +\hat{d}^{\dagger}_{\downarrow}(0) . \nonumber
\end{eqnarray}
From equations Eq.~(\ref{A.2},~\ref{A.3}) we obtain
\begin{eqnarray} \label{A.8}
\hat{d}_{\uparrow}(s) &=& \frac{M^{(-)}_{\downarrow}(s) \hat{A}(s)-iK(s)\hat{B}(s)}{ M^{(+)}_{\uparrow}(s)
M^{(-)}_{\downarrow}(s)+ K^{2}(s) } .
\end{eqnarray}
In order to find $\hat{d}_{\downarrow}(s)$ one should repeat the above described procedure for corresponding set of
equations of motion for operators: $\hat{d}_{\downarrow}$, $\hat{d}_{\uparrow}^{\dagger}$, $\hat{c}_{{\bf k}
\downarrow}$, $\hat{c}_{{\bf k} \uparrow}^{\dagger}$, $\hat{c}_{{\bf q} \downarrow}$, $\hat{c}_{-{\bf q}
\uparrow}^{\dagger}$, $\hat{c}_{{\bf q} \uparrow}^{\dagger}$ and $\hat{c}_{-{\bf q} \downarrow}^{\dagger}(t)$,
respectively. In result we have:
\begin{eqnarray} \label{A.9}
\hat{d}_{\downarrow}(s) &=& \frac{M^{(-)}_{\uparrow}(s) \hat{B}^{\dagger}(s)+iK(s)\hat{A}^{\dagger}(s)}{
M^{(+)}_{\downarrow}(s) M^{(-)}_{\uparrow}(s)+K^2(s) } .
\end{eqnarray}
Additionally, also the expression for $\hat{c}_{{\bf q} \sigma}$ required  in calculations of the current flowing
between the QD and the superconducting lead can be obtained from the set of Eqs~(\ref{EOM.A}-\ref{EOM.H})
\begin{eqnarray}
&& \hat{c}_{{\bf q}\sigma}(s) = \frac{1}{s^{2}+\varepsilon_{\bf q}^{2} +\left| \Delta \right|^{2}} \left( -i V_{\bf q}
(s-i\varepsilon_{\bf q}) \hat{d}_{\sigma}(s) \right. \label{A.10} \\ & & +  \left. \alpha \Delta V_{\bf q}
\hat{d}_{-{\sigma}}^{\dagger}(s) - i\alpha\Delta \hat{c}^{\dagger}_{-{\bf q} - {\sigma}}(0) +(s-i\varepsilon_{\bf
q})\hat{c}_{{\bf q}\sigma}(0) \right) \nonumber ,
\end{eqnarray}
where $\alpha=+ (-)$ for $\sigma=\uparrow (\downarrow)$. The Laplace transforms of $\hat{d}^{\dagger}_{\sigma}(s)$  can
be obtained taking the hermitian conjugation  of the corresponding operator $\hat{d}_{\sigma}(s)$,
Eqs.~(\ref{A.8},\ref{A.9}). Note, that in the wide-band-limit approximation the functions $M^{(\pm)}_{\sigma}(s)$ and
$K(s)$ can be expressed in the superconducting atomic limit $\Delta=\infty$ in the forms $s\pm i
\varepsilon_{\sigma}+\Gamma_{N}/2$ and $\Gamma_{S}/2$, respectively. Finally, as an example, we give here in the
explicit form the Laplace transform of  $\hat{d}_{\uparrow}(t)$:
\begin{eqnarray}
&& \hat{d}_{\uparrow}(s) = \frac{1}{(s-s_{3})(s-s_{4})} \left\{ \left( s -
i\varepsilon_{\downarrow}+\frac{\Gamma_{N}}{2}\right) \right. \label{A.11} \\ && \times \left[ \hat{d}_{\uparrow}(0) -
i\sum_{\bf k} \frac{V_{\bf k} \; \hat{c}_{{\bf k}\uparrow}(0)} {s+i\varepsilon_{\bf k}} - i \sum_{\bf q} \frac{V_{\bf
q}(s-i\varepsilon _{\bf q})\hat{c}_{{\bf q}\uparrow}(0)}{s^{2}+\varepsilon^{2}_{\bf q}+\Delta^{2}} \right. \nonumber
\\ && \left. - \sum_{\bf q} \frac{V_{\bf q}\Delta \hat{c}_{-{\bf q}\downarrow}^{\dagger}(0)}
{s^{2}+\varepsilon^{2}_{\bf q}+\Delta^{2}}  \right] -i \frac{\Gamma_{S}}{2} \left[  \hat{d}_{\downarrow}^{\dagger}(0) +
i\sum_{\bf k} \frac{V_{\bf k} \; \hat{c}_{{\bf k}\downarrow}^{\dagger}(0)} {s-i\varepsilon_{\bf k}} \right. \nonumber
\\ && \left. \left. + i \sum_{\bf q} \frac{V_{\bf q}(s+i\varepsilon_{\bf q})\hat{c}_{{\bf q}
\downarrow}^{\dagger}(0)}{s^{2}+\varepsilon^{2}_{\bf q}+\Delta^{2}} + \sum_{\bf q} \frac{V_{\bf q}\Delta \hat{c}_{-{\bf
q}\uparrow}(0)} {s^{2}+\varepsilon^{2}_{\bf q}+\Delta^{2}} \right] \right\} . \nonumber
\end{eqnarray}
The expression for $\hat{d}_{\uparrow}^{\dagger}(s)$ can be obtained taking the hermitian conjugation of
$\hat{d}_{\uparrow}(s)$ and making the replacement  $(s_3,s_4)\leftrightarrow (s_1,s_2)$, where
\begin{eqnarray}
s_{1,2} & = & \frac{1}{2} \left[ i(\varepsilon_{\uparrow}-\varepsilon_{\downarrow}) -\Gamma_{N}\pm i \sqrt{\delta}
\right] ,
\label{A.12} \\
s_{3,4} & = & \frac{1}{2} \left[- i(\varepsilon_{\uparrow}-\varepsilon_{\downarrow}) -\Gamma_{N}\pm i \sqrt{\delta}
\right] , \nonumber\\
 \delta & = & (\varepsilon_{\uparrow}+\varepsilon_{\downarrow})^2 + \Gamma_S^2 . \nonumber
\end{eqnarray}
Note, that  in the formula for $\hat{d}_{\sigma}(s)$  there is still present the superconduction energy gap parameter 
($\Delta$) in all operator terms. The limit $\Delta \rightarrow \infty$ will be done later in calculations of the average
values of the product of two corresponding operators, e.g. $\left<
\hat{d}_{\sigma}^{\dagger}(t)\hat{d}_{\sigma}(t)\right>$ or $\left< \hat{d}_{\sigma}^{\dagger}(t)\hat{c}_{{\bf
k}'\sigma}(t)\right>$.

%
\subsection{QD occupancy\label{sec:LaplaceA2}} \label{A2}
We calculate the QD occupancy, $n_{\sigma}(t)$, according to the formula Eq.~\ref{QD_occupancy} taking the expectation
value of the product of two corresponding inverse Laplace transforms. As at $t=0$ the QD is decoupled from the external
reservoirs, the only nonvanishing expectation values  would comprise the following averages $\left<
\hat{d}_{\sigma}^{\dagger}(0)\hat{d}_{\sigma}(0)\right> = n_{\sigma}(0)$, $\left<
\hat{d}_{\sigma}(0)\hat{d}_{\sigma}^{\dagger}(0)\right> = 1- n_{\sigma}(0)$, $\left< \hat{c}_{{\bf
k}\sigma}^{\dagger}(0)\hat{c}_{{\bf k}'\sigma}(0)\right> = \delta_{{\bf k},{\bf k}'} f_{N}(\varepsilon_{\bf k})$, and
$\left< \hat{c}_{{\bf k}\sigma}(0)\hat{c}_{{\bf k}'\sigma}^{\dagger}(0)\right> = \delta_{{\bf k},{\bf k}'} \left[ 1 -
f_{N}(\varepsilon_{\bf k}) \right]$, where $f_{N}(\varepsilon_{\bf k})$ is the Fermi-Dirac distribution of the normal
lead. Other terms, corresponding to  mobile electrons of the superconducting lead, can be neglected in the limit
$\Delta=\infty$ (we return to this problem later), but they can be of course included when considering the finite
energy gap. Finally, using Eq.~(\ref{A.8}), the QD occupancy can be expressed in the form:
\begin{widetext}
\begin{eqnarray}
n_{\sigma}(t) &=& {\cal{L}}^{-1} \left\{ \frac{s+i\varepsilon_{-\sigma}+\frac{\Gamma_{N}}{2}}{(s-s_1)(s-s_2)}
\right\}(t){\cal{L}}^{-1} \left\{ \frac{s-i\varepsilon_{-\sigma}+\frac{\Gamma_{N}}{2}} {(s-s_3)(s-s_4)} \right\}(t) \;
\left< \hat{d}_{\sigma}^{\dagger}(0)\hat{d}_{\sigma}(0)\right>
\label{A.14} \\
&+& \left( \frac{\Gamma_{S}}{2} \right)^{2} {\cal{L}}^{-1} \left\{ \frac{1}{(s-s_1)(s-s_2)} \right\}(t){\cal{L}}^{-1}
\left\{ \frac{1}{(s-s_3)(s-s_4)} \right\}(t) \;
\left< \hat{d}_{-\sigma}(0)\hat{d}_{-\sigma}^{\dagger}(0)\right> \nonumber \\
&+& \left( \frac{\Gamma_{S}}{2} \right)^{2} \sum_{{\bf k},{\bf k}'} V_{\bf k}V_{\bf k'} {\cal{L}}^{-1} \left\{ \frac{1}
{(s-s_1)(s-s_2)(s+i\varepsilon_{\bf k})}\right\}(t) {\cal{L}}^{-1} \left\{ \frac{1}{(s-s_3)(s-s_4)(s-i\varepsilon_{{\bf
k}'})} \right\}(t) \left< \hat{c}_{{\bf k}-\sigma}(0) \hat{c}_{{\bf
k}'-\sigma}^{\dagger}(0)\right> \nonumber\\
&+&  \sum_{{\bf k},{\bf k}'} V_{\bf k}V_{\bf k'} {\cal{L}}^{-1} \left\{
\frac{s+i\varepsilon_{-\sigma}+\frac{\Gamma_{N}}{2}} {(s-s_1)(s-s_2)(s-i\varepsilon_{\bf k}) } \right\}(t)
{\cal{L}}^{-1} \left\{ \frac{s-i\varepsilon_{-\sigma}+\frac{\Gamma_{N}}{2}}{(s-s_3)(s-s_4)(s+i\varepsilon_{{\bf k}'}) }
\right\}(t) \left< \hat{c}_{{\bf k}\sigma}^{\dagger}(0) \hat{c}_{{\bf k}'\sigma}(0) \right> \nonumber ,
\end{eqnarray}
where for $\sigma=\downarrow$ the replacement $(s_1,s_2,s_3,s_4) \rightarrow (s_3,s_4,s_1,s_2)$ should be made. In the
wide-band limit approximation we can recast the third and fourth terms to the form:
\begin{eqnarray}
\frac{\Gamma_{S}^2}{4} \frac{\Gamma_{N}}{2\pi} \int_{-\infty}^{\infty} d\varepsilon \left[ 1 - f_{N} (\varepsilon)
\right] {\cal{L}}^{-1} \left\{ \frac{1}{(s-s_1)(s-s_2)(s+i\varepsilon)}
\right\}(t) {\cal{L}}^{-1} \left\{ \frac{1}{(s-s_3) (s-s_4)(s-i\varepsilon)} \right\}(t) \label{A.16}\\
+ \frac{\Gamma_{N}}{2\pi} \int_{-\infty}^{\infty} d\varepsilon f_{N} (\varepsilon) {\cal{L}}^{-1} \left\{
\frac{s+i\varepsilon_{-\sigma}+\frac{\Gamma_{N}} {2}}{(s-s_1)(s-s_2)(s-i\varepsilon)} \right\}(t) {\cal{L}}^{-1}
\left\{ \frac{s-i\varepsilon_{-\sigma}+\frac{\Gamma_{N}} {2}}{(s-s_3) (s-s_4)(s+i\varepsilon)} \right\}(t) . \nonumber
\end{eqnarray}
\end{widetext}
The final explicit formula for $n_{\sigma}(t)$ is somewhat lengthy, therefore we present it here for the case
$\varepsilon_{\sigma}=0$:
\begin{eqnarray}
n_{\sigma}(t) &=& n_{\sigma}(0)e^{-\Gamma_{N} t} + \left(1- n_{-\sigma}(0)\right) e^{-\Gamma_{N} t} \; \mbox{\rm
sin}^{2} \left( \frac{\Gamma_S}{2}\;t\right) \nonumber \\
& + &  \frac{\Gamma_{N}}{2\pi} \int_{-\infty}^{\infty} \hspace{-0.3cm} d\varepsilon \; f_N(\varepsilon) \;
F_1(\varepsilon,t) \label{A.17}  \\
& + &  \frac{\Gamma_{N}}{2\pi} \frac{\Gamma_S^2}{4}  \int_{-\infty}^{\infty} \hspace{-0.3cm} d\varepsilon \; \left[ 1 -
f_N(\varepsilon) \right] F_2(\varepsilon,t)  . \nonumber
\end{eqnarray}
Functions $F_1(\varepsilon,t)$ and $F_2(\varepsilon,t)$ have the following analytical forms:
\begin{eqnarray}
F_1(\varepsilon,t) &=& \frac{1}{A(\varepsilon)} \left\{ \frac{e^{-\Gamma_N t}}{2} \left[
\left(\frac{\Gamma_{N}^2}{4}-\frac{\Gamma_{S}^2}{4}+\varepsilon^2 \right)\cos(\Gamma_S
  t) \right. \right. \nonumber\\
  &-& \left. \frac{\Gamma_{N}\Gamma_S}{2} \sin(\Gamma_S t)+\frac{\Gamma_{N}^2}{4}+\frac{\Gamma_{S}^2}{4}+\varepsilon^2
  \right] \label{A.18}\\
&-& e^{-\Gamma_N t/2} \left[ 2 \left(\frac{\Gamma_{N}^2}{4}+\varepsilon^2 \right)\cos(\varepsilon t) \cos\left(
\frac{\Gamma_S}{2}\;t\right)
\right. \nonumber\\
&-& \frac{\Gamma_{N}\Gamma_S}{2} \cos(\varepsilon t) \sin\left( \frac{\Gamma_S}{2}\;t\right) \nonumber\\
&+& \left. \left. \Gamma_S \varepsilon \sin(\varepsilon t) \sin\left( \frac{\Gamma_S}{2}\;t\right) \right] +
\frac{\Gamma_{N}^2}{4}+\varepsilon^2 \right\} , \nonumber
\end{eqnarray}
\begin{eqnarray}
F_2(\varepsilon,t) &=& \frac{1}{\Gamma_S A(\varepsilon)} \left\{ e^{-\Gamma_N t} \left[ \frac{-2}{\Gamma_S}
\left(\frac{\Gamma_{N}^2}{4}-\frac{\Gamma_{S}^2}{4}+\varepsilon^2 \right)\cos(\Gamma_S t) \right. \right. \nonumber\\
&+& \left. \Gamma_{N} \sin(\Gamma_S t)+\frac{2}{\Gamma_{S}} \left(
\frac{\Gamma_{N}^2}{4}+\frac{\Gamma_{S}^2}{4}+\varepsilon^2 \right) \right] \label{A.19}\\
&+& e^{-\Gamma_N t/2} \left[ 2 \left(\varepsilon_{-} \cos(\varepsilon_+ t)-  \varepsilon_{+} \cos(\varepsilon_- t)
\right) \right.
\nonumber\\
&-& \left. \left. \Gamma_N \left( \sin(\varepsilon_+ t) -  \sin(\varepsilon_- t)\right) \right] + \Gamma_S \right\} ,
\nonumber
\end{eqnarray}
and $A(\varepsilon)=\left(\frac{\Gamma_{N}^2}{4}+\varepsilon_-^2 \right) \left(\frac{\Gamma_{N}^2}{4}+\varepsilon_+^2
\right)$, $\varepsilon_{+/-}=\varepsilon \pm \frac{\Gamma_S}{2}$. It should be noted that for $\Gamma_S=0$ the formula
(\ref{A.14}) for $n_{\sigma}(t)$ reduces to the standard expression obtained by the non-equilibrium Green's
function technique, e.g. \cite{Jauho-1994}:
\begin{eqnarray}
n_{\sigma}(t) &=& n_{\sigma}(0)e^{-\Gamma_{N} t} +  \frac{\Gamma_{N}}{\pi} e^{-\Gamma_{N} t/2} \int_{-\infty}^{\infty}
\hspace{-0.3cm}
d\varepsilon \; f_N(\varepsilon)  \nonumber  \\
& \times & \frac{\cosh(\Gamma_N t/2)-\cos((\varepsilon-\varepsilon_{\sigma})t)
}{\frac{\Gamma_{N}^2}{4}+(\varepsilon-\varepsilon_{\sigma})^2} , \label{A.20}
\end{eqnarray}
Let us return to the discussion about the terms which appear in general formula, Eq.~(\ref{QD_occupancy}), but involve
operators $\hat{c}_{{\bf q} \sigma}$. Let us analyze one of these terms e.g.
\begin{eqnarray}
 \left< {\cal{L}}^{-1} \left\{ \frac{s+i\varepsilon_{\downarrow}+\frac{\Gamma_{N}}{2}}{(s-s_1)(s-s_2)} \sum_{\bf q}
\frac{V_{\bf q}(s+i\varepsilon_{\bf q}) \hat{c}_{{\bf q} \uparrow}^{\dagger}(0)}{s^2+\varepsilon_{\bf q}^2 + \Delta^2} \right\}(t) \right. \nonumber\\
\left. \cdot {\cal{L}}^{-1} \left\{ \frac{s-i\varepsilon_{\downarrow}+\frac{\Gamma_{N}}{2}}{(s-s_3)(s-s_4)} \sum_{{\bf
q}'} \frac{V_{{\bf q}'}(s-i\varepsilon_{{\bf q}'}) \hat{c}_{{\bf q}' \uparrow}(0) }{s^2 +\varepsilon_{{\bf q}'}^2 +
\Delta^2}  \right\}(t) \right> \,,\nonumber\\
\end{eqnarray}
which can be reduced  to the form:
\begin{eqnarray}
&& \frac{\Gamma_S}{2\pi} \int_{-\infty}^{+\infty} d\varepsilon f_S(\varepsilon) {\cal{L}}^{-1} \left\{ \frac{\left(
s+i\varepsilon_{\downarrow}+\frac{\Gamma_{N}}{2}\right)\left(s+i\varepsilon \right)}{(s-s_1)(s-s_2)(s^2+\varepsilon^2 +
\Delta^2) } \right\}(t)  \nonumber\\ && {\cal{L}}^{-1} \left\{ \frac{\left(
s-i\varepsilon_{\downarrow}+\frac{\Gamma_{N}}{2}\right)\left(s-i\varepsilon \right)}{(s-s_3)(s-s_4)(s^2+\varepsilon^2 +
\Delta^2) } \right\}(t)  ,
\end{eqnarray}
where we have used the equality $ \left< \hat{c}_{{\bf q} \uparrow}^{\dagger}(0) \hat{c}_{{\bf q}' \uparrow}(0) \right>
= \delta_{{\bf q}{\bf q}'}f_s(\varepsilon_{{\bf q}})$. We have checked numerically (integrating the product of two
corresponding inverse Laplace transforms) that this integral is smaller and smaller with increasing $\Delta$, so for
$\Delta = \infty$ we put it equal to zero. Similarly, we have also checked all other terms involving operators
$\hat{c}_{{\bf q} \sigma}(0)$ and found they can be neglected for $\Delta = \infty$.

\begin{widetext}
\subsection{QD pair correlation function\label{sec:LaplaceA3}} \label{A3}

Using the formulas for $\hat{d}_{\sigma}(s)$, Eq.~(\ref{A.8},\ref{A.9}) and performing similar calculations 
as for the QD occupancy, the induced on-dot pairing, $\left< \hat{d}_{\downarrow}(t) \hat{d}_{\uparrow}(t)\right>$,  
can be written in the general form as:
\begin{eqnarray}
\chi(t) &=& i \frac{\Gamma_{S}}{2} \; \left[ n_{\uparrow}(0)  \; {\cal{L}}^{-1} \left\{ \frac{1}{(s-s_1)(s-s_2)}
\right\}(t) {\cal{L}}^{-1} \left\{ \frac{s-i\varepsilon_{\downarrow}+\frac{\Gamma_{N}}{2}} {(s-s_3)(s-s_4)} \right\}(t)
\right.
\label{A.23} \\
&-&  \left( 1 - n_{\downarrow}(0) \right) \;
 {\cal{L}}^{-1} \left\{ \frac{s-i\varepsilon_{\uparrow}+\frac{\Gamma_{N}}{2}}
{(s-s_1)(s-s_2)} \right\}(t) {\cal{L}}^{-1} \left\{ \frac{1}{(s-s_3)(s-s_4)} \right\}(t)
\nonumber \\
&-& \frac{\Gamma_{N}}{2\pi} \int_{-\infty}^{\infty} d\omega \left[ 1 - f_{N} (\omega) \right] {\cal{L}}^{-1} \left\{
\frac{s-i\varepsilon_{\uparrow}+\frac{\Gamma_{N}} {2}}{(s-s_1)(s-s_2)(s+i\omega)}  \right\}(t) {\cal{L}}^{-1} \left\{
\frac{1}{(s-s_3) (s-s_4)(s-i\omega)} \right\}(t)
\nonumber \\
&+& \left. \frac{\Gamma_{N}}{2\pi} \int_{-\infty}^{\infty} d\omega f_{N}(\omega) {\cal{L}}^{-1} \left\{
\frac{1}{(s-s_1)(s-s_2)(s-i\omega)}  \right\}(t) {\cal{L}}^{-1} \left\{
\frac{s-i\varepsilon_{\downarrow}+\frac{\Gamma_{N}}{2}}{(s-s_3)(s-s_4)(s+i\omega)} \right\}(t) \right] \; . \nonumber
\end{eqnarray}
\end{widetext}
%
%
\subsection{QD-superconducting lead current  \label{sec:LaplaceA4}} \label{A4}

Starting with the formula similar to the one given in Eq.~(\ref{eqn12}) the QD-superconductor 
current can be obtained from
\begin{eqnarray}
j_{S\sigma}(t)=2\mbox{\rm Im} \sum_{\bf q} V_{\bf q} \left< {\cal{L}}^{-1} \left\{
\hat{d}_{\sigma}^{\dagger}(s)\right\}(t) {\cal{L}}^{-1} \left\{ \hat{c}_{{\bf q}\sigma}(s)\right\}(t) \right>
\nonumber\\ \label{A.25}
\end{eqnarray}
where $\hat{d}_{\sigma}^{\dagger}(s)$ and $\hat{c}_{{\bf q}\sigma}(s)$ are given in Eqs.~(\ref{A.8}) and (\ref{A.10}).
Performing similar calculations as in previous subsections let us consider the non-vanishing term proportional to
$n_{\uparrow}(0)$. It has the form
\begin{eqnarray}
&& 2 n_{\uparrow}(0) \mbox{\rm Im} \left\{ -i\sum_{\bf q} V_{\bf q} {\cal{L}}^{-1} \left\{ \frac{
s+i\varepsilon_{\downarrow}+\frac{\Gamma_{N}}{2}}{(s-s_1)(s-s_2)} \right\}(t) \label{A.26} \right. \nonumber\\ \\
&&  \times \left[ {\cal{L}}^{-1} \left\{ \frac{V_{\bf q}\left( s-i\varepsilon_q\right)
\left(s-i\varepsilon_{\downarrow}+ \frac{\Gamma_{N}}{2} \right)}{(s-s_3)(s-s_4)(s^2+\varepsilon_{\bf q}^2 + \Delta^2) }
\right\}(t) \right. \nonumber\\ &+& \left. \left. \frac{\Gamma_S}{2} {\cal{L}}^{-1} \left\{ \frac{V_{\bf q}
\Delta}{(s-s_3)(s-s_4)(s^2+\varepsilon_{\bf q}^2 + \Delta^2) } \right\}(t) \right] \right\} . \nonumber
\end{eqnarray}
The first part of this equation vanishes for $\Delta=\infty$ and the second part we calculate interchanging the
summation over ${\bf q}$ with the Laplace transformation. In result we have:
\begin{eqnarray}
&& 2 n_{\uparrow}(0) \mbox{\rm Im} \left[ -i \frac{\Gamma_S}{2} {\cal{L}}^{-1} \left\{ \frac{
s+i\varepsilon_{\downarrow}+\frac{\Gamma_{N}}{2}}{(s-s_1)(s-s_2)} \right\}(t) \right. \\ && \left. {\cal{L}}^{-1}
\left\{\frac{1}{(s-s_3)(s-s_4)} \sum_{\bf q} \frac{V_{\bf q}^2 \Delta}{(s^2+\varepsilon_{\bf q}^2 + \Delta^2) }
\right\}(t) \right] . \label{A.27} \nonumber
\end{eqnarray}
As $\sum_{\bf q} \frac{V_{\bf q}^2 \Delta}{(s^2+\varepsilon_{\bf q}^2 + \Delta^2) }=\frac{\Gamma_S}{2}$ for
$\Delta=\infty$, then finally the term proportional to $n_{\uparrow}(0)$ takes the form:
\begin{eqnarray}
&& 2 n_{\uparrow}(0) \frac{\Gamma_S^2}{4} \mbox{\rm Im} \left[ -i  {\cal{L}}^{-1} \left\{ \frac{
s+i\varepsilon_{\downarrow}+\frac{\Gamma_{N}}{2}}{(s-s_1)(s-s_2)} \right\}(t) \right. \nonumber\\ && \left.
{\cal{L}}^{-1} \left\{\frac{1}{(s-s_3)(s-s_4)} \right\}(t) \right] .  \label{A.28}
\end{eqnarray}
In similar way we calculate the terms proportional to $\left<\hat{d}_{\uparrow}(0) \hat{d}_{\uparrow}^{\dagger}(0)
\right>$, $\left<\hat{c}_{{\bf k}\uparrow}^{\dagger}(0) \hat{c}_{{\bf k}' \uparrow}(0) \right>$  and
$\left<\hat{c}_{{\bf k}\uparrow}(0) \hat{c}_{{\bf k}' \uparrow}^{\dagger}(0) \right>$. All other terms containing the
expectation values of two superconducting lead electron operators vanish (for $\Delta=\infty$), similarly as in the
case for $n_{\uparrow}(t)$. Finally, we get the general equation for $j_{S \sigma}(t)$ 
\begin{widetext}
\begin{eqnarray}
j_{S\sigma}(t) &=&  \frac{\Gamma^2_{S}}{2}  \mbox{\rm Re} \left[-n_{\sigma}(0)  \; {\cal{L}}^{-1} \left\{
\frac{s+i\varepsilon_{-\sigma}+\frac{\Gamma_{N}}{2}} {(s-s_1)(s-s_2)} \right\}(t)  {\cal{L}}^{-1} \left\{ \frac{1}
{(s-s_3)(s-s_4)} \right\}(t)
\right. \label{current_j_S} \\
&+& \left. \left( 1 - n_{-\sigma}(0) \right) \;
 {\cal{L}}^{-1} \left\{ \frac{1}{(s-s_1)(s-s_2)} \right\}(t)  {\cal{L}}^{-1} \left\{
\frac{s+i\varepsilon_{\sigma}+\frac{\Gamma_{N}}{2}}{(s-s_3)(s-s_4)} \right\}(t)  + \frac{\Gamma_N}{2\pi} \Phi_{\sigma}
\right] \,, \nonumber
\end{eqnarray}
where
\begin{eqnarray} \label{A.277}
\Phi_{\sigma} &=& - \int_{-\infty}^{\infty} d\varepsilon f_{N}(\varepsilon)
 {\cal{L}}^{-1} \left\{ \frac{s+i\varepsilon_{-\sigma}+\frac{\Gamma_{N}}
{2}}{(s-s_1)(s-s_2)(s-i\varepsilon)}  \right\}(t) {\cal{L}}^{-1} \left\{ \frac{1}{(s-s_3) (s-s_4)(s+i\varepsilon)}
\right\}(t)
 \\
&+& \int_{-\infty}^{\infty} d\varepsilon \left[ 1 - f_{N} (\varepsilon) \right] {\cal{L}}^{-1} \left\{
\frac{1}{(s-s_1)(s-s_2)(s+i\varepsilon)}  \right\}(t) {\cal{L}}^{-1} \left\{
\frac{s+i\varepsilon_{\sigma}+\frac{\Gamma_{N}}{2}}{(s-s_3)(s-s_4)(s-i\varepsilon)} \right\}(t)  . \nonumber
\end{eqnarray}
%
%
For $\sigma=\downarrow$ the replacement $(s_1, s_2, s_3, s_4) \rightarrow (s_3, s_4, s_1, s_2)$ should be done.
Note that for $\mu_N=0$  the formula for the superconducting current simplifies as $\mbox{\rm Re} \Phi_{\sigma}=0$. To
show this property we assume the case $\varepsilon_{\sigma}=0$ and express $\Phi_{\sigma}$ in the following form:
\begin{eqnarray} \label{A.278}
\Phi_{\sigma} &=& \sum_{j=1}^4 \int_{-\infty}^{\infty} d\varepsilon (1-2 f_{N}(\varepsilon)) A_j(\varepsilon) \,,
\end{eqnarray}
where
\begin{eqnarray}
A_1(\varepsilon) &=&  \frac{-i e^{-\Gamma_N t}}{2\Gamma_S}  \left[ \frac{e^{i\Gamma_S t}}
{\left(\frac{\Gamma_N}{2}-i\varepsilon_+\right)\left(\frac{\Gamma_N}{2}+i\varepsilon_-\right)} \right.
 \nonumber\\
&-& \left. \frac{e^{-i\Gamma_S
t}}{\left(\frac{\Gamma_N}{2}-i\varepsilon_-\right)\left(\frac{\Gamma_N}{2}+i\varepsilon_+\right)} \right] ,
 \nonumber\\
A_2(\varepsilon) &=&  \frac{i e^{-\Gamma_N t}
\varepsilon}{\left(\frac{\Gamma_N^2}{4}+\varepsilon_+^2\right)\left(\frac{\Gamma_N^2}{4}+\varepsilon_-^2\right)} ,
 \nonumber\\
A_3(\varepsilon) &=& - \frac{\frac{\Gamma_N}{2}+i\varepsilon}
{\left(\frac{\Gamma_N^2}{4}+\varepsilon_-^2\right)\left(\frac{\Gamma_N^2}{4}+\varepsilon_+^2\right)} \,, \nonumber
\label{A.24}
\end{eqnarray}
%
%
\begin{eqnarray}
 A_4(\varepsilon) &=& -\frac{ e^{-\Gamma_N t/2}}{2}  \left[ \frac{e^{-i\varepsilon_+ t}}
{\left(\frac{\Gamma_N^2}{4}+\varepsilon_+^2\right)\left(\frac{\Gamma_N}{2}-i\varepsilon_-\right)} +
\frac{e^{-i\varepsilon_- t}}
{\left(\frac{\Gamma_N^2}{4}+\varepsilon_-^2\right)\left(\frac{\Gamma_N}{2}-i\varepsilon_+\right)} \right]
 \nonumber\\
&-&  \frac{i e^{-\Gamma_N t/2}}{\Gamma_{S}}\left(\frac{\Gamma_N}{2}+i\varepsilon \right)  \left[
\frac{-e^{i\varepsilon_+ t}}
{\left(\frac{\Gamma_N^2}{4}+\varepsilon_+^2\right)\left(\frac{\Gamma_N}{2}+i\varepsilon_-\right)} +
\frac{e^{i\varepsilon_- t}}
{\left(\frac{\Gamma_N^2}{4}+\varepsilon_-^2\right)\left(\frac{\Gamma_N}{2}+i\varepsilon_+\right)} \right] \,,
 \nonumber\\
\end{eqnarray}
\end{widetext}
and  $\varepsilon_{+/-}=\varepsilon \pm \frac{\Gamma_S}{2}$. For $\mu_N=0$ and zero temperature the function
$\Phi_{\sigma}$ can be written in the form:
\begin{eqnarray} \label{A.279}
\Phi_{\sigma} &=& \sum_{j=1}^4 \int_{0}^{\infty} d\varepsilon \left( A_j(\varepsilon)-A_j(-\varepsilon)\right) \,,
\end{eqnarray}
and using the following properties of $A(\varepsilon)$ functions, $A_1(\varepsilon)=A_1(-\varepsilon)$,
$A_2(\varepsilon)=-A_2(-\varepsilon)$, $\mbox{\rm Re} A_3(\varepsilon)=\mbox{\rm Re} A_3(-\varepsilon)$ and
$A_4(-\varepsilon)=A_4^*(\varepsilon)$, we find that $\mbox{\rm Re} \Phi_{\sigma}=0$. This conclusion is also valid for
$\varepsilon_{\sigma}\neq 0$.

\section{Mean field approximation\label{sec:HFB}}
\label{B}

Let us consider the effective Hamiltonian of the proximitized QD coupled to the normal lead, treating correlations
within the Hartree-Fock-Bogoliubov approximation
\begin{eqnarray}
\hat{H}&=&\sum_{\sigma} \left( \varepsilon_{\sigma}+Un_{-\sigma}(t)\right)
\hat{d}_{\sigma}^{\dagger} \hat{d}_{\sigma}
+ \left[ \left( \frac{\Gamma_{S}}{2} + U \chi(t) \right) \hat{d}_{\uparrow}^{\dagger}
\hat{d}_{\downarrow}^{\dagger}  \right. \nonumber \\
&+& \left. \mbox{\rm h.c.} \right]
+ \sum_{{\bf k},\sigma} \left[ V_{\bf k} \hat{d}_{\sigma}^{\dagger}  \hat{c}_{{\bf k} \sigma}
+ \mbox{\rm h.c.} \right] +
\sum_{{\bf k},\sigma} \varepsilon_{{\bf k}\sigma} \hat{c}_{{\bf k} \sigma}^{\dagger}
\hat{c}_{{\bf k} \sigma} .
\end{eqnarray}
In general, all parameters $\varepsilon_{\sigma}$, $\Gamma_{S}$, $V_{\bf k}$, $\varepsilon_{{\bf k}\sigma}$ can be
time-dependent. We outline the algorithm for numerical computation of the QD charge $n_{\sigma}(t)$ and the induced
pairing $\chi(t)=\left< \hat{d}_{\downarrow}(t)\hat{d}_{\uparrow}(t)\right>$. We have to solve numerically the
following set of coupled equations of motion 
%
\begin{eqnarray}
\frac{d n_{\sigma}(t)}{dt} &=& 2 \mbox{\rm Im} \left(  \sum_{\bf k} V_{\bf k} e^{i\varepsilon_{\vec k}t}\left<
\hat{d}_{\sigma}^{\dagger}(t) \hat{c}_{{\bf k}\sigma}(0) \right>
\right. \nonumber \\
&-& \left. \bar{\Delta}^{*}(t) \chi(t) - \Gamma_{N} n_{\sigma}(t) \right) \,, \label{B.1}
\\
\frac{d \chi(t)}{dt} &=& i \sum_{\bf k} V_{\bf k} e^{-i\varepsilon_{\bf k}t} \left( \left< \hat{d}_{\uparrow}(t)
\hat{c}_{{\bf k}\downarrow}(0)\right> - \left< \hat{d}_{\downarrow}(t) \hat{c}_{{\bf k}\uparrow}(0)\right> \right)
\nonumber \\
& - & \left[ i \left( \bar{\varepsilon}_{\uparrow}(t) +
\bar{\varepsilon}_{\downarrow}(t) \right) + \Gamma_{N} \right] \chi(t)
\nonumber \\
&-& i \bar{\Delta}(t) \left( 1 - n_{\downarrow}(t) - n_{\uparrow}(t) \right) \,, \label{B.2}
\end{eqnarray}
where $\bar{\varepsilon}_{\sigma}(t)=\varepsilon_{\sigma}+U n_{-\sigma}(t)$ and
$\bar{\Delta}(t)=\frac{\Gamma_{S}}{2}+U\chi(t)$. Here we have used the wide-band-limit approximation and assumed
$\varepsilon_{{\bf k}\sigma}$ to be time-independent. The new functions appearing in (\ref{B.1},\ref{B.2}) can be
determined solving corresponding equations of motion
\begin{eqnarray}
&&\frac{d}{dt} \left< \hat{d}_{\sigma}^{\dagger}(t)\hat{c}_{{\bf k}\sigma}(0)\right>
= \left( i \bar{\varepsilon}_{\sigma}(t)-\frac{\Gamma_{N}}{2}\right)
\left< \hat{d}_{\sigma}^{\dagger}(t)\hat{c}_{{\bf k}\sigma}(0)\right>
\nonumber \\
&& +i\alpha \bar{\Delta}^{*}(t) \left< \hat{d}_{-\sigma}(t)\hat{c}_{{\bf k}\sigma}(0)\right>
+ iV_{\bf k} e^{i\varepsilon_{\bf k}t} f_{N}(\varepsilon_{\bf k})
\label{B.3} \\
&&\frac{d}{dt} \left< \hat{d}_{\sigma}(t)\hat{c}_{{\bf k}-\sigma}(0)\right>
= - \left( i \bar{\varepsilon}_{\sigma}(t)+\frac{\Gamma_{N}}{2}\right)
\nonumber \\
&& \times \left< \hat{d}_{\sigma}(t)\hat{c}_{{\bf k}-\sigma}(0)\right>
-i\alpha \bar{\Delta}(t) \left< \hat{d}_{\sigma}^{\dagger}(t)\hat{c}_{{\bf k}\sigma}(0)\right> ,
\label{B.4}
\end{eqnarray}
where $\alpha=+(-)$ for $\sigma=\uparrow (\downarrow)$ and
$f_{N}(\varepsilon_{\bf k})$ is the Fermi-Dirac distribution of
mobile electrons in the normal lead.

\section{Subgap Kondo effect\label{sec:Kondo}}
\label{C}

When the Coulomb potential $U$ is sufficiently large in comparison to $\Gamma_{S}$ the QD ground state evolves towards
the spinful (doublet) configuration $\left| \sigma \right>$. Under such conditions the effective spin exchange between
the correlated QD and mobile electrons of the metallic lead activate the subgap Kondo effect. It has been analyzed by
many groups, using various techniques \cite{Rodero-11}.  In the present context we shall make use of basic facts,
pointed out recently by R.\ \v{Z}itko et al \cite{Zitko-2015} and independently by one of us
\cite{Domanski-2016,Zonda-2016}.

The exchange interaction $-\sum_{{\bf k},{\bf p}}J_{{\bf k,p}}\;\;\hat{{\bf S}}_{d} \cdot \hat{{\bf S}}_{{\bf k}{\bf
p}}$ between the QD spin $\hat{{\bf S}}_{d}$ and spins  $\hat{{\bf S}}_{{\bf k}{\bf p}}$ of the mobile electrons in
normal lead can be determined by means of the generalizing canonical Schrieffer-Wolff transformation. Adopting it  to
the N-QD-S setup it has been found, that for the superconducting atomic limit the exchange coupling near the Fermi
energy $J_{{\bf k}_{F},{\bf k}_{F}}$ is equal to \cite{Domanski-2016}
\begin{eqnarray}
J_{{\bf k}_{F},{\bf k}_{F}}=\frac{U \left|V_{{\bf k}_{F}} \right|^{2}}{\varepsilon_{\sigma}\left(\varepsilon_{\sigma}+U
\right) +(\Gamma_{S}/2)^{2}} . \label{generalized_SW}
\end{eqnarray}
For a spinful configuration the Kondo temperature can be estimated e.g.\ using the Bethe-Ansatz formula  $T_{K}\propto
\mbox{{\rm exp}} \left\{ -1/\left[2\rho(\varepsilon_{F})J_{{\bf k}_{F}{\bf k}_{F}} \right]\right\}$, where
$\rho(\varepsilon_{F}$) is the density of states of the normal lead at the Fermi level. We have compared such results
with the unbiased NRG calculations and it has been found that the Kondo temperature is expressed by
\cite{Domanski-2016}
\begin{eqnarray}
T_{K}=\eta\;\frac{\sqrt{\Gamma_{N}U}}{2}\;{\rm exp}\!
\left[\pi\;\frac{\varepsilon_{\sigma}\left(\varepsilon_{\sigma}+U\right)
+(\Gamma_{S}/2)^{2}}{\Gamma_{N}U}\right]\label{T_K}
\end{eqnarray}
with $\eta\approx 0.6$. In particular, for the half-filled quantum dot ($\varepsilon_{\sigma}=-U/2$) the exchange
couling (\ref{generalized_SW}) simplifies to
\begin{equation}
J_{{\bf k}_{F},{\bf k}_{F}} = J_{{\bf k}_{F},{\bf k}_{F}}^{(N)} \; \frac{U^{2}}{U^{2}-\Gamma_{S}^{2}} ,
\label{effective_T_K}
\end{equation}
where $J_{{\bf k}_{F},{\bf k}_{F}}^{(N)}$ stands for the normal case ($\Gamma_{S}\!=\!0$). Upon approaching a
transition from the spinful doublet to the BCS-like (spinless) ground state the Kondo temperature is substantially
enhanced \cite{Zitko-2015,Domanski-2016}
\begin{equation}
T_{K} =  T_{K}^{(N)} \; \mbox{\rm exp} \left[ \frac{\pi}{\Gamma_{N}U} \left( \frac{\Gamma_{S}}{2} \right)^{2} \right] .
\label{effective_T_K}
\end{equation}

\begin{figure}
\includegraphics[width=0.95\columnwidth]{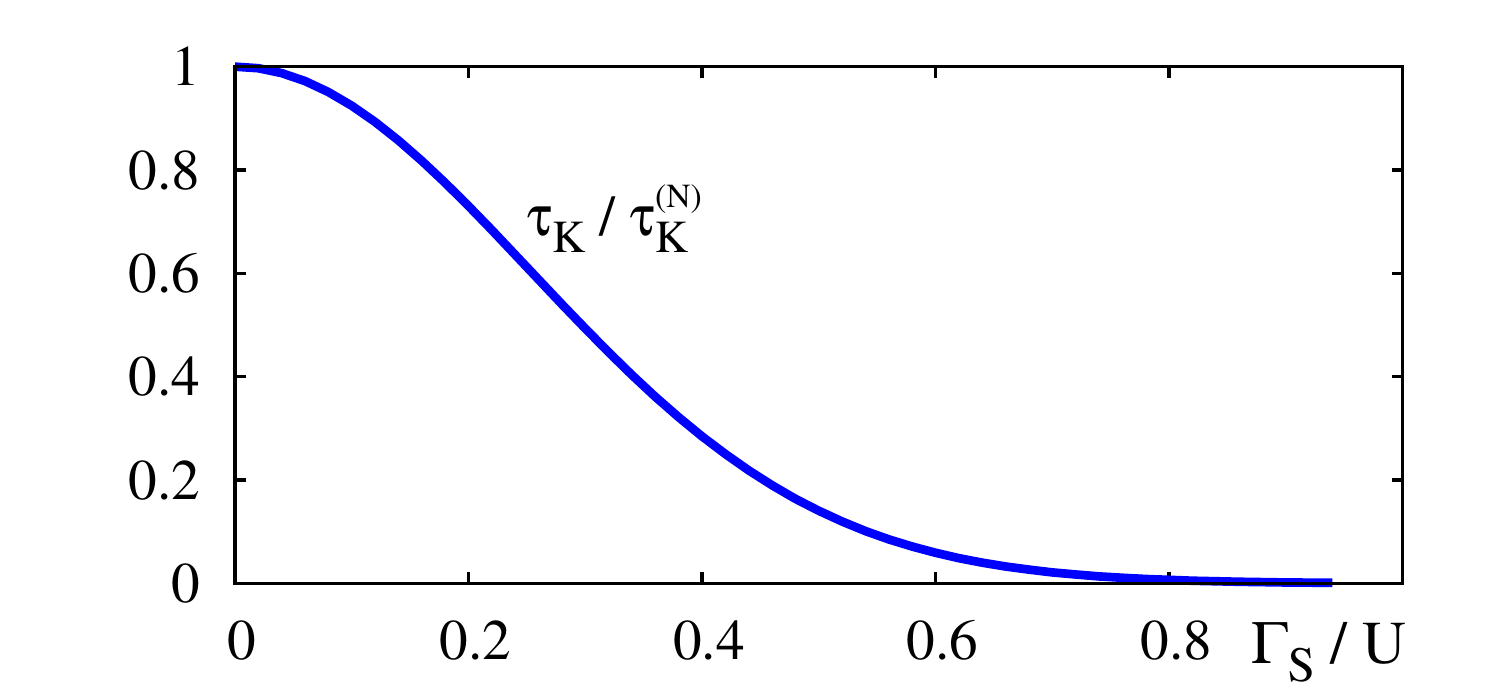}
\caption{Characteristic time-scale $\tau_{K} \sim 1/T_{K}$ of the subgap Kondo effect obtained for the half-filled QD
using $U=10\Gamma_{N}$ with respect to varrying ratio $\Gamma_{S}/U$. In this case the spinful ground state exists in
the region  $\Gamma_{S}\in (0, U)$.} \label{Kondo}
\end{figure}

To get some insight into the transient phenomena related with the subgap Kondo regime we  make use of the final
conclusions inferred in Ref.\ \cite{Nordlander-99} from the time-dependent noncrossing approximation study. The
characteristic time $\tau_{K}$ needed for the Abrikosov-Suhl peak to emerge at the Fermi level has been found  to scale
inversely with the Kondo temperatutre, i.e.\ $\tau_{K}\sim 1/T_{K}$. This information adopted to our N-QD-S setup
implies the following relative ratio for the half-filled QD
\begin{equation}
\tau_{K} =  \tau_{K}^{(N)} \; \mbox{\rm exp} \left[ - \frac{\pi}{\Gamma_{N}U} \left( \frac{\Gamma_{S}}{2} \right)^{2}
\right] , \label{Kondo_time}
\end{equation}
where $\tau_{K}^{(N)}$ stands for the normal state value ($\Gamma_{S}=0$). We plot this scaling in Fig.\ \ref{Kondo}.
Let us remark, that many-body screening (\ref{generalized_SW}) of the QD spin can be practically realized only in the
doublet ground state (which for the half-filled QD occurs when $\Gamma_{S}<U$). By increasing the ratio $\Gamma_{S}/U$
the Andreev bound states tend to their crossing and simultaneously the Abrikosov-Suhl peak (\ref{effective_T_K})
quickly broadens \cite{Zitko-2015,Domanski-2016}. This explains why the characteristic time-scale $\tau_{K}$ strongly
decreases with respect to $\Gamma_{S}/U$. More systematic analysis of this phenomenon is beyond a scope of the present
paper.

\bibliography{biblio}

\end{document}